\documentclass{aa}
\usepackage{natbib}
\bibpunct{(}{)}{;}{a}{}{,}
\usepackage{txfonts}
\usepackage{graphicx}

\title{Radiative Transfer Modeling of Three-Dimensional \\ Clumpy AGN Tori and its Application to NGC\,1068}

\author{S.~F.~H\"onig \and
T.~Beckert \and
K.~Ohnaka \and
G.~Weigelt}
\institute{Max-Planck-Institut f\"ur Radioastronomie, Auf dem H\"ugel 69, 53121 Bonn, Germany}

\offprints{S.~F.~H\"onig, \\ \email{shoenig@mpifr-bonn.mpg.de}}
\date{Received 2 December 2005 / Accepted 22 February 2006}

\abstract{Recent observations of NGC\,1068 and other AGN support the idea of a geometrically and optically thick dust torus surrounding the central supermassive black hole and accretion disk of AGN. In type 2 AGN, the torus is seen roughly edge-on, leading to obscuration of the central radiation source and a silicate absorption feature near $10\,{\rm\mu m}$. While most of the current torus models distribute the dust smoothly, there is growing evidence that the dust must be arranged in clouds. We describe a new method for modeling near- and mid-infrared emission of 3-dimensional clumpy tori using Monte Carlo simulations. We calculate the radiation fields of individual clouds at various distances from the AGN and distribute these clouds within the torus region. The properties of the individual clouds and their distribution within the torus are determined from a theoretical approach of self-gravitating clouds close to the shear limit in a gravitational potential. We demonstrate that clumpiness in AGN tori can overcome the problem of over-pronounced silicate features. Finally, we present model calculations for the prototypical Seyfert 2 galaxy NGC\,1068 and compare them to recent high-resolution measurements. Our model is able to reproduce both the SED and the interferometric observations of NGC\,1068 in the near- and mid-infrared.
		}

\authorrunning{S.~F.~H\"onig et al.}
\titlerunning{3D Radiative Transfer Modeling of Clumpy Tori and its Application to NGC\,1068}

\begin{document}

\maketitle

\section{Introduction} \label{sec1}

The {\it Unification Scheme} for AGN was introduced by \citet{Mil83} to explain the different appearances of various types of AGN. According to our present understanding, the centers of all AGN host a central black hole of several $10^6$ to $10^{10}\,\mathrm{M_{\sun}}$. It is surrounded by an accretion disk which reaches inwards to the closest stable Keplerian orbit. Due to energy gained from accretion in a gravitational potential, the accretion disk is heated up to $\sim$$10^5\,\mathrm{K}$, which results in strong emission in the UV/optical part of the spectrum. Such a UV-bump has been observed in QSOs as well as in Seyfert 1 galaxies.

At distances where the ambient temperature falls below $\sim\!10^3\,{\rm K}$, dust is able to survive; it is thus possible to have a large dusty structure present which surrounds the AGN. This dust reservoir can explain the optical properties of type 1 and type 2 AGN. Type 1 AGN show strong emission in the blue and UV regime of the spectrum, indicating a direct view to the accretion disk. Additionally, broad emission lines are present, which come from fast-moving material close to the central black hole. Type 2 objects do not show the blue/UV-bump. Emission lines are narrow, indicating that the material responsible for the emission is moving much slower and, thus, must be further away from the center. The difference between type 1 and type 2 AGN is explained by obscuration of the central region. The obscuring structure is presumably an optically thick dusty torus, probably starting at the dust sublimation radius $r_{\rm subl}$ and reaching out to $10$ to $10^2\,r_{\rm subl}$ \citep[e.g.,][]{Gra94,Boc00,Rad03}. In type~1 AGN this torus is viewed approximately pole-on, while in type 2 objects it is seen close to edge-on. This implies that most type 2 objects harbor a type 1 nucleus -- a suggestion which was confirmed by the observations of weak polarized broad emission lines hidden by strong narrow lines in the Seyfert 2 galaxy NGC\,1068 \citep{Mil83}. The observed ratio of Seyfert 1 to Seyfert 2 galaxies of approximately 1 to 1.3 \citep{Mai95,Lac04} suggests that the torus is both optically and geometrically thick.

Additional evidence for the existence of a dust torus comes from the silicate feature at $\sim$$10\,{\rm \mu m}$ in the spectral energy distribution (SED) of AGN. In type 1 AGN, this feature is expected to be detected in emission since the hot dust at the inner rim of the torus can be seen directly. Recent mid-infrared (MIR) spectra obtained with instruments on board the Spitzer satellite confirm the silicate emission feature in QSOs \citep{Sie05,Hao05} and the LINER NGC\,3998 \citep{Stu05}. For type 2 objects, the dust feature is observed in absorption \citep[e.g.,][]{Jaf04} due to obscuration by the cold dust. Further evidence of a torus comes from near-infrared (NIR) bispectrum speckle interferometry \citep{Wit98,Wei04} and speckle interferometry \citep{Wei99} of the resolved nucleus of NGC\,1068. $H$- and $K^\prime$-band images \citep{Wei04} show a resolved core structure of $1.3\,\times\,2.8\,{\rm pc}$. This resolved core is interpreted as dust close to the sublimation radius and an inner outflow cavity. Furthermore, strong evidence for a dust torus in NGC\,1068 arises from long-baseline interferometry of NGC\,1068 with the Very Large Telescope Interferometer (VLTI). \citet{Jaf04} report on $8-13\,{\rm\mu m}$ MIR observations with the VLTI MIDI instrument. The wavelength dependence of the visibility inside and outside the silicate absorption feature can be interpreted with a two-component model consisting of a hot (800\,K) dust structure with an extension of $<1\,{\rm pc}$ and a warm (320\,K) structure of $3.4\times\,2.1\,{\rm pc}$. Finally, first near-infrared $K$-band long-baseline interferometry of NGC\,1068 with the VLTI and the VINCI instrument was reported by \citet{Wit04}. These VINCI observations detected a substructure with a size of $\le 3\,{\rm mas}$ ($\la 0.2\,{\rm pc}$).

First pioneering radiative transfer modeling of AGN dust tori was carried out by \citet{Pie92,Pie93}. They studied a torus with cylindrical geometry and a homogeneous smooth dust distribution. Further investigations with smooth dust distributions were reported by \citet{Gra94}, \citet{Efs95}, and \citet{Gra97}. Recently, \citet{Sch05} presented very detailed radiative transfer modeling, introducing a physically motivated geometry and dust distribution \citep{Cam95}.

\citet{Kro88} discussed that smooth dust distributions probably cannot survive in the vicinity of an AGN. They proposed that the dust is arranged in clouds and that the observed velocity dispersion is actually a result of their orbital motion within the gravitational potential. First results of radiative transfer calculations of such a clumpy medium were reported by \citet{Nen02} and \citet{Dul05}. They show that a torus consisting of clouds is capable of suppressing the strength of silicate emission and absorption, as has been observed \citep[e.g.,][]{Jaf04,Sie05,Hao05,Stu05}. It also became clear that the actual cloud distribution is a critical point. Therefore, physically motivated models of the torus are required.

In this paper we present our 3D Monte Carlo radiative transfer modeling of clumpy AGN tori. We have developed a method which allows us to obtain high-resolution Monte Carlo simulations with relatively small computation times. We calculate the radiation fields of individual clouds at various distances from the AGN and distribute these clouds within the torus region. After discussing the principle of our method, we present model spectra and images of clumpy tori and analyze the influence of various parameters. To illustrate the feasibility of our approach, we finally model the SED and visibilities of NGC\,1068 simultaneously.

\section{Monte Carlo simulation of individual dust clouds}

This section gives an overview of the Monte Carlo radiative transfer treatment with our code {\sf mcsim\_mpi} \citep{Ohn05} and describes its application to dust clouds. In the following, we present cloud SEDs and analyze some aspects of the Monte Carlo treatment.

\subsection{AGN primary radiation} \label{AGNrad}

\begin{figure}
\centering
\includegraphics[angle=0,width=8.5cm]{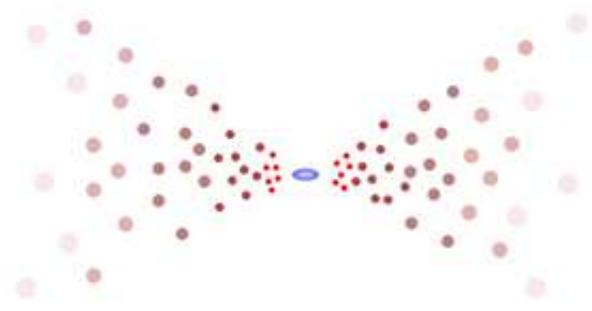}
\caption{Sketch of a clumpy torus. The hot and cold clouds are randomly distributed around a central accretion disk (blue). The size and optical thickness varies with radial distance.}\label{Princ}
\end{figure}

We consider a central super-massive black hole (smBH) which is fed by an accretion disk. The smBH and its accretion disk are surrounded by an optically and geometrically thick dust torus. \citet{Kro88} argue that the temperature has to be $\sim$$10^6\,{\rm K}$ if the velocity despersion is of the order of $100\,{\rm km\,s^{-1}}$. This temperature greatly exceeds the dust sublimation temperature. Thus, we arrange the dust in clouds. In Fig. \ref{Princ}, we illustrate the clumpy composition of the torus.

\begin{figure}
\centering
\includegraphics[angle=0,width=8.5cm]{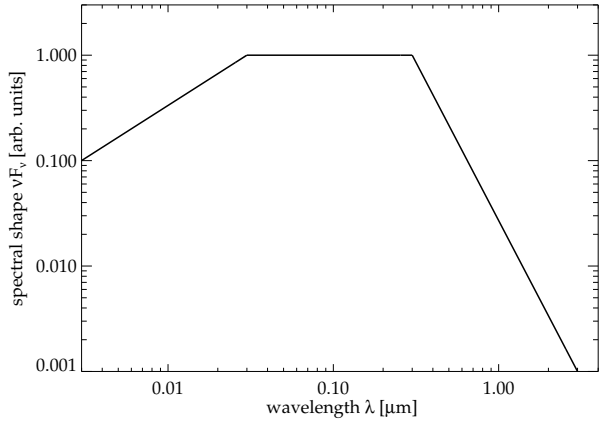}
\caption{AGN source spectrum as used in our Monte Carlo simulations.}\label{fig1}
\end{figure}

The primary source of radiation is the accretion disk. We approximate the SED of the accretion disk by a broken power law:
\begin{equation}
\lambda F_{\lambda} \propto \left\{ \begin{array}{lcl}
                         \lambda & & \lambda < 0.03\,{\rm\mu m} \\ \nonumber
                         {\rm constant} & & 0.03\,{\rm\mu m} \le \lambda \le 0.3\,{\rm\mu m} \\ \nonumber
                         \lambda^{-3} & & \lambda > 0.3\,{\rm\mu m} \nonumber
                          \end{array} \right.
\end{equation}
It is illustrated in Fig.~\ref{fig1}. The SED is a fit to QSO spectra \citep{Man98} with a characteristic steep decline towards the MIR.

\subsection{Geometry and dust cloud properties}\label{dust}

Our radiative transfer code {\sf mcsim\_mpi} allows the modeling of dust distributions of arbitrary 3-dimensional geometry. Since there is no knowledge about the shape of dust clouds around AGN, we model spherically symmetric clouds. For our calculations, we assume a standard galactic dust composition of 53\% astronomical silicates and 47\% graphite \citep{Dra84}. The absorption and scattering efficiencies $Q_{\rm abs}$ and $Q_{\rm sca}$ for MRN distributions of grain sizes \citep{Mat77} have been calculated with the {\sf DUSTY} code \citep{Ive99}. In our Monte Carlo code, we use an average grain size of $0.1\,{\rm\mu m}$ representing the whole distribution \citep[e.g.][]{Efs94,Wol03,Nen05}.

For the characterization of the cloud radiation field, three main parameters are needed: the bolometric AGN luminosity $L_{\rm AGN}$, the cloud's optical depth $\tau_{cl}$, and its distance $r$ from the central source. $\tau_{cl}$ is taken at the reference visual wavelength of $0.55\,{\rm \mu m}$ and denotes the optical depth through the cloud center. At other wavelengths, the optical depth is derived from $Q_{\rm abs}$. Within a cloud, the dust is distributed homogeneously.

\subsection{Radiative Transfer Code}

Our radiative transfer code {\sf mcsim\_mpi} is based on the Monte Carlo approach by \citet{Bjo01}. A detailed description of our code is given in \citet{Ohn05}. Here we briefly summarize the concept. As for any Monte Carlo approach, the luminosity of the central radiation source is divided into monochromatic photon packages. If a photon package hits a dust particle, it is either scattered or absorbed. In case the photon package is scattered, a random scattering angle is determined according to the differential cross section. If it is absorbed, the dust particle is heated up and the photon is re-emitted at a frequency determined by its temperature. The approach by Bjorkman \& Wood has the advantage that it is non-iterative for fixed opacities and thus requires less computing time. As output, the code provides SEDs for scattered and absorbed photons, which are then added to a complete cloud SED. Furthermore, it is possible to obtain a number of cloud SEDs for different observer positions (=phase angles $\phi$).

\begin{figure}
\centering
\includegraphics[angle=0,width=8.5cm]{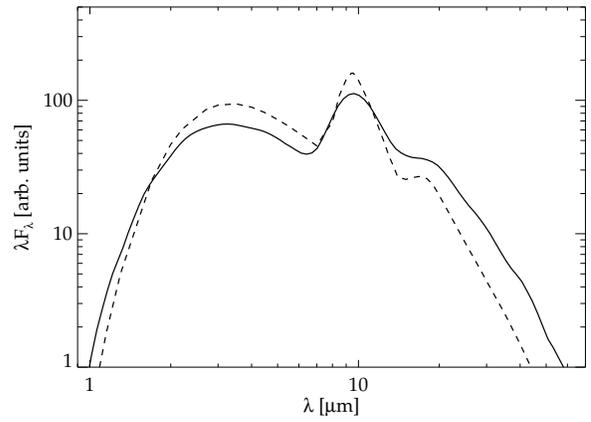}
\caption{Comparison of a hot-side cloud SED simulated with our Monte Carlo code {\sf mcsim\_mpi} (solid line) and a SED of a dust sphere with external isotropic illumination (obtained with the {\sf DUSTY} code; dashed line). The Monte Carlo calculations were performed with $5\times10^8$ photon packages (see Sect. \ref{QCMCS}).}\label{fig3}
\end{figure}

The code has been tested intensively and was recently applied to the modeling of VLTI MIDI visibilities of the evolved star IRAS\,08002$-$3803 \citep{Ohn05}. For testing our specific cloud geometry, we compared our cloud SEDs to results obtained with the {\sf DUSTY} code. {\sf DUSTY} allows the modeling of a dust sphere in an external isotropic radiation field. Although this is not exactly the same geometry as our external illumination by a central source, we use the resulting SEDs for comparison with the hot-side SEDs of our {\sf mcsim\_mpi} clouds. While the overall SED shapes are quite similar (see Fig.~\ref{fig3}), the resulting differences are expected due to different illumination patterns.
\begin{figure}
\centering
\includegraphics[angle=0,width=6.5cm]{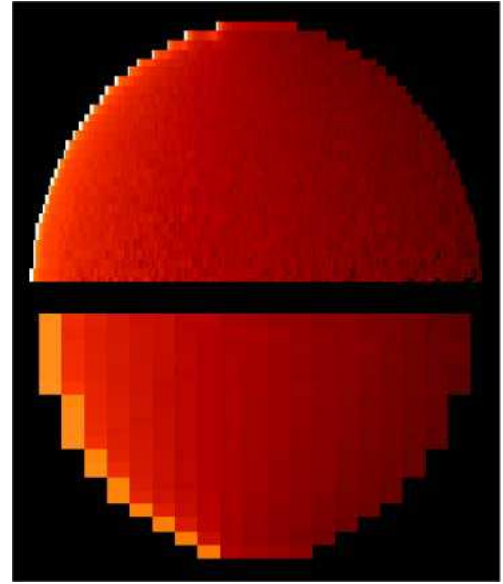}
\caption{Temperature profile of a simulated cloud ($\tau_{cl}=150$) with a high-resolution grid ($150\times32$ cells; top) and a low-resolution grid ($20\times20$ cells; bottom). The number of photon packages are $5\times10^8$ for both simulations. The cloud was placed close to the dust sublimation radius, where the AGN is to the left. The hottest temperatures are $\sim$1400\,K (white), the coolest are $\sim$200\,K (dark red).}\label{figGrid}
\end{figure}

\subsection{Quality checks of Monte Carlo simulations}\label{QCMCS}

\begin{table}
\caption{Temperature gradient variations for increasing numbers of radial cells within a cloud ($T_{\rm hot}=1500\,{\rm K}$).}\label{tab1}
\centering
\vspace{0.1cm}
\begin{tabular}{r c r c r}
\hline\hline
number & & $\Delta T(0.1\,R_{cl})$ & & $\Delta T(R_{cl})$ \\
radial cells & & [K] & & [K] \\ \hline
10 & & 560 & & 660 \\
20 & & 580 & & 800 \\
40 & & 600 & & 980 \\
70 & & 610 & & 1090 \\
100 & & 640 & & 1160 \\
150 & & 640 & & 1180 \\ \hline
\end{tabular}
\end{table}

\begin{figure*}
\centering
\includegraphics[angle=0,width=16.5cm]{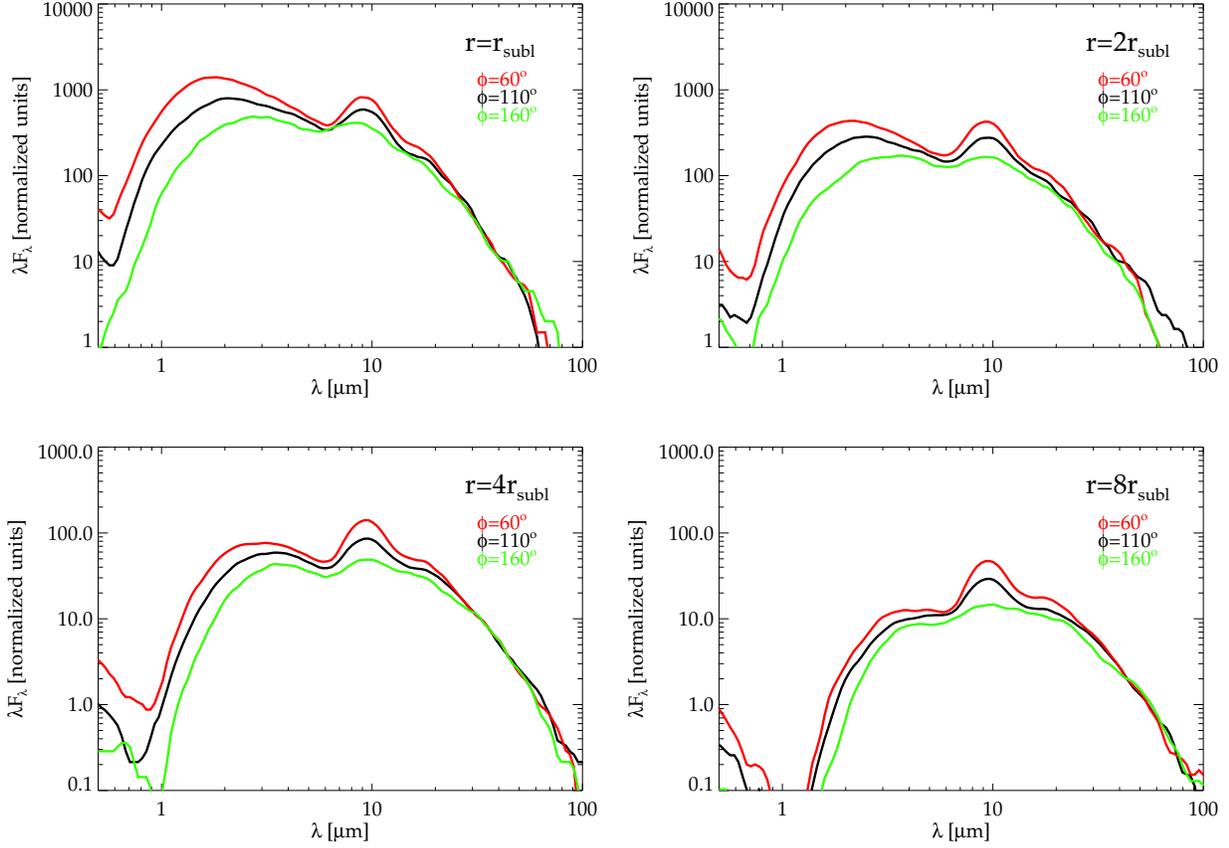}
\caption{SEDs of spherical dust clouds with optical depths of $\tau_{cl}=70$ simulated with our Monte Carlo code {\sf mcsim\_mpi}. In the upper row, the panels represent clouds at distances $r=r_{\rm subl}$ and $r=2\, r_{\rm subl}$; in the lower row they represent clouds at distances $r=4\, r_{\rm subl}$ and $r=8\, r_{\rm subl}$. For each cloud three SEDs are shown, as seen under phase angles of $\phi=160\degr$ (green), $\phi=110\degr$ (black), and $\phi=60\degr$ (red). In terms of illumination, these phase angles correspond to 3\%, 33\%, and 75\%, respectively. The absolute flux values are normalized to the cloud surfaces.}\label{fig4}
\end{figure*}

Depending on $\tau_{cl}$, the temperature within a cloud decreases from the hot to the cold side. If the number of grid cells is too small, this temperature profile will not be reproduced properly and parts of the cloud are either too cold or too hot. This specifically applies to clouds with high $\tau_{cl}$ since the temperature gradients can be steep. In Table~\ref{tab1} we show Monte Carlo temperature gradients for a cloud with $\tau_{cl}=100$ at a distance of $r=r_{\rm subl}$ ($T(r_{\rm subl})=1500\,{\rm K}$) from the AGN. We list the temperature gradient from the hot side to one tenth of the cloud diameter inward, as well as the complete gradient from the front to the back side. As can be seen, the temperature gradients converge once a certain number of radial cells is used. Therefore, it is quite important to use enough radial cells to sample the temperature profile within a wide range of $\tau_{cl}$ values. In Fig.~\ref{figGrid} we show the comparison of a cloud simulated with high and low resolution. For both simulations we used $5\times10^8$ photon packages (see next paragraph). In the low-resolution simulation, the temperatures on the hot side are too cold compared to the high-resolution simulation.

Furthermore, we have to ensure that enough photon packages interact in each grid cell. As discussed above, we use 100--150 grid cells in radial direction. Since the temperature gradients in the plane perpendicular to the radial direction do not need a very high resolution, we use only 32 cells. The 3-dimensional radiation field results from the rotational symmetry around the central radial axis. Thus, our clouds consist of $\sim\,5\times 10^3$ grid cells. Tests with the code show that the results are quite good for $10^6-10^7$ photon packages interacting in the cloud grid cells. If we take less than $\sim\,10^6$ photon packages, the resulting SED becomes too noisy.

\subsection{Cloud SEDs} \label{clsed}

We performed calculations for a number of different cloud parameters to study their SEDs and consequences for a clumpy torus. For the simulations we used $10^9$ photon packages and $\sim5\times10^3$ grid cells for each cloud.

In Fig.~\ref{fig4} we show our results for clouds with $\tau_{cl}=70$. The clouds were placed at different distances from the AGN (1, 2, 4, and $8\, r_{\rm subl}$). For each cloud we show SEDs for three different phase angles $\phi$. The phase angle is defined as the angle between the direction from the cloud to the observer and the direction from the cloud to the AGN, as seen from the cloud. $\phi=0\degr$ corresponds to a cloud that appears fully illuminated, while $\phi=180\degr$ represents the backside view of the cloud without direct AGN illumination. This concept is similar to the phases of the moon: depending on where the cloud is located around the AGN, only a fraction of its surface seen by the observer is directly illuminated by the AGN.

The different panels in Fig.~\ref{fig4} show the dependence of the continuum peak on the distance to the AGN. The peak emission of the SED continuum shifts to longer wavelengths for larger distances. In addition, the absolute flux decreases. At larger distances from the AGN, the relative flux of the silicate emission at $\sim$$10\,{\rm\mu m}$ increases with respect to the continuum. Furthermore, it is remarkable that the silicate emission feature is not as pronounced as in torus model SEDs of smooth dust distributions. Although the SEDs of Fig~\ref{fig4} are model SEDs of individual clouds, the implications for a torus spectrum are evident: because of the clumpy structure, the emission feature will not become very strong (see Sect.~\ref{Results} and \ref{ModSEDN1068}).

\begin{figure}
\centering
\includegraphics[angle=0,width=8.5cm]{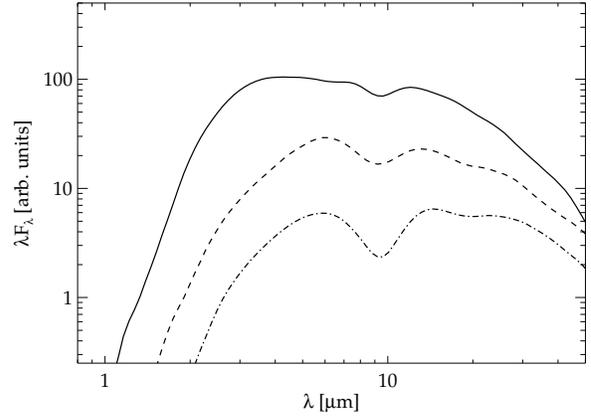}
\caption{Model SEDs of the unilluminated side of dust clouds at two different distances from the central source. The silicate feature is in absorption. The solid and the dashed SEDs represent clouds at $2\, r_{\rm subl}$ with  $\tau_{cl}=100$ and $\tau_{cl}=400$, respectively. The dashed-dotted SED comes from a cloud at $4\, r_{\rm subl}$ and $\tau_{cl}=250$.}\label{fig5}
\end{figure}

It is interesting to follow the SED development from the hot frontside to the cold backside view of each cloud. The cloud with the smallest distance to the AGN (cf. upper left panel in Fig.~\ref{fig4}) shows a clear shift of the continuum peak from around $1.7\,{\rm\mu m}$ on the hot side to around $3\,{\rm\mu m}$ on the cold side. In addition, the silicate feature in emission starts to decrease with respect to the continuum. This is a result of the temperature gradient within the cloud. On the hot side, the temperature is high enough to produce an emission feature. The emission basically takes place in the upper dust layers of the hot parts. When the cloud is seen from the cold side, the hot side's emission has to pass through colder layers within the cloud and suffers from absorption. Furthermore, the colder layers produce less or no silicate emission. As a result, the overall feature decreases. At larger distances (cf. lower right panel in Fig.~\ref{fig4}), the emission from the hot side barely compensates for the absorption within the cloud, so the silicate emission disappears in the SED of the cloud's cold side. In extreme cases of high $\tau_{cl}$, large distances $r$, or no direct illumination, the emission may even turn into absorption. This effect is illustrated in Fig.~\ref{fig5}, where SEDs from the cold side of clouds are shown for such cases.

\section{Torus assembly}

We will now present a simple parametrization of the torus. In the following, we describe our method for arranging the clouds within the torus. This method allows us to obtain model SEDs and model images of tori using the simulated cloud SEDs which have been calculated before. Afterwards, a physical model for the cloud properties is outlined. To distinguish between properties of clouds and the torus, we will label all cloud parameters with the subscript $cl$, while non-subscripted parameters always apply to the torus as a whole.

\subsection{Torus parameters}\label{TorModels}

To assemble the clumpy torus, it is necessary to get information about the cloud properties and the cloud distribution within the torus. A very simple approach to this problem is the modeling of these properties with free parameters and corresponding power laws. Geometric torus input parameters are the total number of clouds within the torus $N_{\rm tot}$, their distribution $\eta_r(r)$ and $\eta_{\varphi}(\varphi)$ in radial and azimuthal direction, respectively, the scale height of the torus $H(r)$, which also determines the cloud distribution in $z$-direction $\eta_z(r,z)$, and its outer rim $r_{\rm o}$ \citep[also see][]{Nen05}.
\begin{table*}
\caption{Parameters of the {\it free parameter model} and the {\it accretion scenario} model.}\label{tab2}
\centering
\vspace{0.1cm}
\begin{tabular}{l c| c | c}
\hline\hline
model parameter & & free parameter model & accretion scenario \\ \hline
number of clouds in equatorial LOV $N_0$ & & free & free \\
optical depth $\tau_{cl}(r)$ & & $\propto r^{\Gamma_\tau}\mathrm{,\,\,}\Gamma_\tau\,\,\mathrm{free}$ & $\propto M(r)^{1/2}r^{-3/2}$ \\
cloud size $R_{cl}(r)$ & & $\propto r^{\Gamma_R}$, $\Gamma_R$ free & $\propto M(r)^{-1/2}r^{3/2}$ \\
cloud distribution $\eta_r(r)$ & & $\propto r^a$, $a$ free & $\propto \left(\frac{M(r)}{\dot{M}}\right)^{1/2}r^{-3/2}$ \\
scale height $H(r)$& & $\propto r^{\Gamma_H}$, $\Gamma_H$ free & $\propto \left(\frac{\dot{M}}{M(r)}\right)^{1/2}r^{3/2}$ \\
outer radius $r_{\rm o}$ & & free & --- \\ \hline
\end{tabular}
\end{table*}

For the radial distribution, we use a power law $\eta_r(r)\propto r^a$. The $z$- and $r$-dependences of $\eta_z$ are parameterized by a Gaussian-like ansatz; i.e., $\eta_z(r,z)=\exp\left(z^2/2H(r)^2\right)$. This accounts for a higher cloud density towards the equatorial plane and flaring of the torus. For simplicity, $\eta_{\varphi}(\varphi)$ is set to unity so that $\eta(r,z,\varphi)=\eta_r(r)\cdot\eta_z(r,z)$ and $\int\eta(r,z=0,\varphi)\,{\rm d}r=1$.

For convenience, we substitute the total number of clouds within the torus $N_{\rm tot}$ by the number of clouds $N_0$ along the line of view to the AGN in the equatorial plane \citep{Nen05}. The relation between $N_{\rm tot}$ and $N_0$ is given by
$$ N_{\rm tot} \propto N_0\cdot\int\frac{\eta(r,z,\varphi)}{R_{cl}^2}\,{\rm d}V\rm{\,.} $$

The basic torus parameters $N_0$, $\eta(r,z,\varphi)$, $H(r)$, and $r_{\rm o}$ are supplemented by the properties of the clouds; in particular, their optical depth $\tau_{cl}(r)$ and radius $R_{cl}(r)$. Both parameters are written as $r$-dependent. Actually, whether $\tau_{cl}$ changes with distance or not is unclear. Theoretical arguments exist for a $\tau_{cl}$ gradient (see Sect. \ref{AccScen}). Henceforth, a simple model with the power law parameters described above and listed in Table~\ref{tab2} is called a {\it free parameter model}.

\subsection{Torus SED compilation and geometric peculiarities}\label{GeomPec}

Once a distribution function for the clouds is selected, random positions for $N_{\rm tot}$ clouds are determined according to $\eta(r,z,\varphi)$. Every position is labeled with the corresponding cloud properties $\tau_{cl}$ and $R_{cl}$, and a template spectrum from the database is associated with this position. As it happens, some of the clouds are not directly illuminated by the AGN since another cloud is along the line of sight to the AGN. Such clouds, however, can be indirectly illuminated by the radiation coming from surrounding clouds. Therefore, we must distinguish between these indirectly illuminated clouds ({\it Second Order Clouds; SOCs}) and clouds that are directly illuminated by the AGN ({\it First Order Clouds; FOCs}).

The treatment of FOCs and SOCs is quite different. For FOCs we simply take the SEDs from the Monte Carlo output, as described in Sect. \ref{clsed}. Since there is no direct AGN illumination for a SOC, one has to approximate the illumination caused by all other clouds. Given the grid resolution that is needed for proper handling of the clumps, the necessary distances between the clouds, and the total number of clouds of several $10^4$, an approximation cannot be avoided since, otherwise, the number of photon packages would be $10^{13}-10^{15}$ and more (see Sect. \ref{QCMCS}). This is difficult to achieve with current computation power.

In general, a SOC is located within the radiation field of the surrounding clouds. To approximate this diffuse radiation, we follow the scheme proposed by \citet{Nen05}. The main contribution to a SOC's ambient radiation comes from the hot side of FOCs in the vincinity of the SOC. If we consider the diffuse radiation field to be isotropic, it is possible to approximate its SED at a given distance $r$ from the AGN by averaging a FOC SED over all phase angles $\phi$. To finally obtain the SED of a SOC, the average FOC spectrum can be taken as the input spectrum for an ambient radiation field in our Monte Carlo code. The stength of the ambient radiation field is derived by taking into account the fraction of the sky covered by FOCs as seen by a SOC near the sublimation radius. Because the ambient radiation field is assumed to be isotropic, the simulated SOC shows an identical SED for any phase angle $\phi$. The results obtained from these calculations form the database for SOC templates. The method is discussed in Appendix~\ref{methdis} in more details.

When all clouds have been associated with template SEDs from the FOC and SOC database, the torus SED can be compiled. For this purpose, a torus inclination angle $i$ has to be chosen. A torus seen pole-on has $i=0\degr$, while the edge-on view corresponds to $i=90\degr$. As predicted by the Unification Scheme, type 1 AGN should have small $i$, whereas $i$ should be larger (e.g., $>45\degr$) for type 2 objects. The inclination $i$ also defines a line of view (LOV) from the observer to the torus clouds. With respect to this LOV, each cloud is seen under a specific phase angle $\phi$, depending on its position in the torus. The corresponding SED as seen by the observer is taken from the templates, and the LOV of each cloud is checked for obscuring clouds along the LOV. If there are other clouds, the template SED is attenuated according to the wavelength-dependent $\tau_{cl}$-profile and the fraction of coverage. The resulting SED is added to the total torus SED. Once all clouds are sifted through, the torus SED is complete. Since this method of compiling the torus spectrum does not demand any special symmetry, it is also applicable to non-axisymmetric cloud distributions or other geometries. Actually, the statistical process of distributing clouds already introduces 3-dimensional asymmetries.

\subsection{Physical torus models and cloud properties}\label{AccScen}

\begin{figure*}
\centering
\includegraphics[angle=0,width=16.5cm]{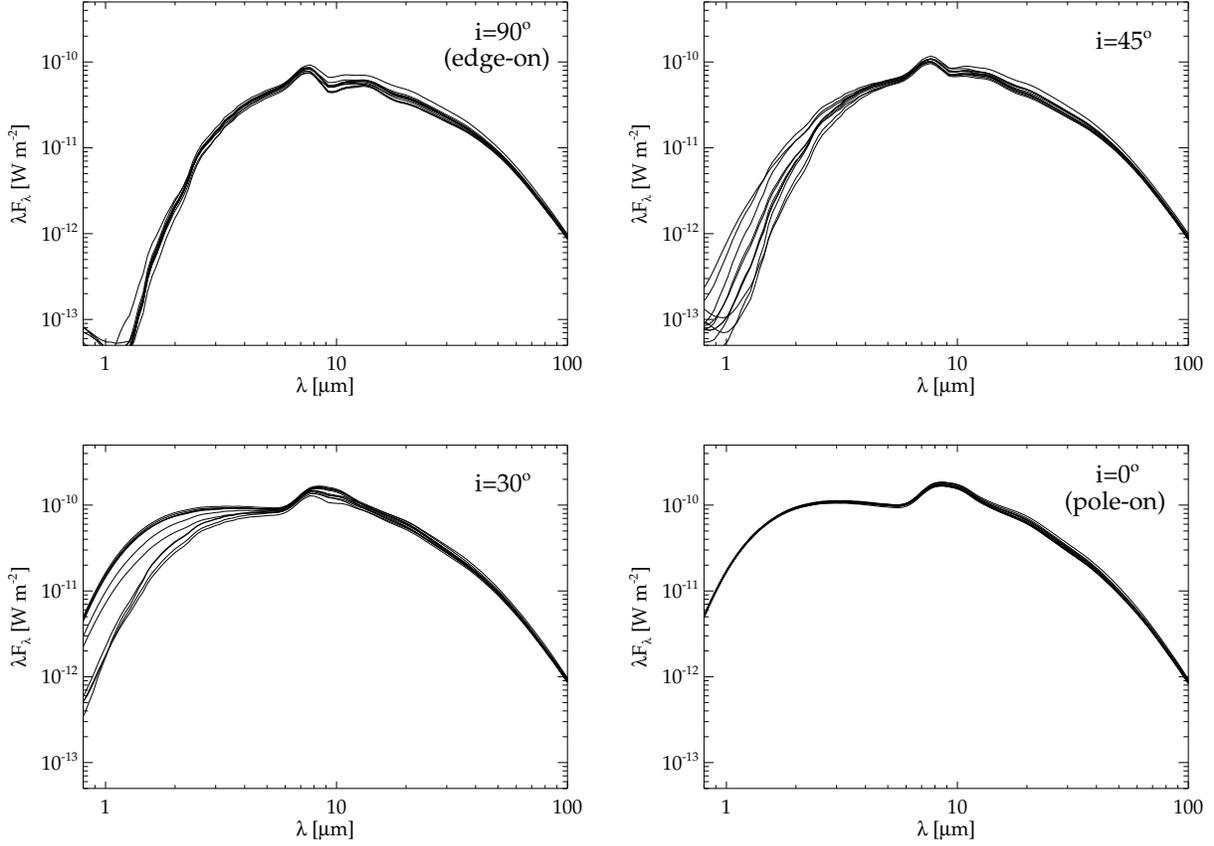}
\caption{Torus SEDs of 10 different random cloud arrangements for the set of parameters listed in Table~\ref{tabparstd}. The panels in the first row show SEDs for inclinations of $i=90\degr$ (edge-on; left) and $i=45\degr$ (right); the panels in the second row show $i=30\degr$ (left) and $i=0\degr$ (pole-on; right).}\label{figVar}
\end{figure*}

Power law models, as described in Sect. \ref{TorModels}, are widely used in torus studies, although it would be desirable to have the parameters fixed by theoretical arguments. For example, \citet{Cam95} describes a smooth dust torus where the dust is produced in stars of a circumnuclear stellar cluster (cnSC). The toroidal geometry is a result of the effective gravitational potential coming from the smBH and the cnSC \citep{Sch05}.

A physically motivated model for {\it clumpy} dust tori was proposed by \citet{Vol04} and \citet{Bec04}. Its basic idea is that the dust clouds within the torus are self-gravitating. Their size is close to the shear limit given by the gravitational potential of the cnSC and the smBH. Actually, the torus is in a region where the influences of both the cnSC and the smBH are important. The cnSC is supposed to be spherically symmetric and isothermal; i.e., the relation of the enclosed mass of the cnSC is given by $M^{\rm SC}(r) \propto r$. In this torus model, the material that forms the torus is accreted from large distances of more than $100\,r_{\rm subl}$.   The {\it accretion scenario} assumes effective angular momentum transfer due to cloud collisions. This leads to large scale heights for high accretion rates above the Eddington limit of the central smBH. Since the enclosed mass in the torus is a combination of smBH and cnSC mass, the Eddington limit within the torus is higher than in the accretion disk. An advantage of this model is that it reduces the number of free torus parameters. In addition to the AGN bolometric luminosity $L_{\rm AGN}$ and its black hole mass $M_{\rm BH}$, one has to specify the mass accretion rate $\dot{M}$ through the torus and the properties of the cnSC. The latter is defined by its core radius $R^{\rm SC}_{\rm core}$ and the core mass $M^{\rm SC}_{\rm core}$. The dependencies of the accretion scenario are listed in Table \ref{tab2}, in comparison to the free parameter model and constraints coming from observations. In the table, $M(r)$ denotes the sum of cnSC and smBH masses. Furthermore, the accretion scenario doesn't need an outer radius for the torus. Since it predicts $R_{cl}\propto r^{3/2}\cdot M(r)^{-1/2}$, and $M_{cl}\propto r^{3/2}\cdot M(r)^{-1/2}$, the optical depth of a cloud is given by
$$\tau_{cl}(r) \propto \frac{M_{cl}}{R_{cl}^2} \propto M(r)^{1/2}\cdot r^{-3/2}\,. $$
For an isothermal cnSC, as observed in the center of our galaxy \citep[e.g.][]{Sch03}, $M(r)$ is proportional to $r$ so that $\tau_{cl}(r)\propto r^{-1}$. This implies that $\tau_{cl}(r)$ steadily decreases with increasing distance from the AGN. At larger distances, the clouds become larger and less dense; they smoothly pass into the surrounding galactic ISM.

Some interesting predictions of the accretion scenario are the cloud sizes and masses. The total mass for clouds -- gas and dust -- close to the sublimation radius $r_{\rm subl}$ should be of the order of $1\ldots5\,{\rm M_{\sun}}$, while the cloud radius is approximately $0.01$ to $0.05\,{\rm pc}$. Since $R_{\rm cl}$ and $M_{\rm cl}$ weakly depend on the enclosed mass, these values are similar for a wide range of smBH and cnSC masses. Henceforth, for the modeling we use cloud properties which are consistent with the accretion scenario.

\section{Results}\label{Results}

In this section we present model SEDs and model images obtained from our torus simulations. The effect of the clumpy structure on the torus is analyzed by parameter studies. We briefly discuss the impact of clumpiness on the modeling of observations. For our parameter studies, we have chosen parameters listed in Table~\ref{tabparstd}. Changes and variations to this set of parameters are noted in the text.

\begin{table}
\caption{Parameters as used in our parameter studies.}\label{tabparstd}
\centering
\vspace{0.1cm}
\begin{tabular}{l l | l l}
\hline\hline
Torus & Model Value & Clouds & Model Value \\ \hline
$L_{\rm AGN}$ & $2\times\,L_{12}$ & $R_{cl}(r)$ & $0.02\,{\rm pc}\times \left(\frac{r}{r_{\rm subl}}\right)^{1.5}$ \\
$r_{\rm subl}$ & $0.8\,{\rm pc}$ & $N_0$ & 8 \\
$d$ & 10\,{\rm Mpc} & $\tau_{cl}$ & $40\times\left(\frac{r}{10\,r_{\rm subl}}\right)^{-0.8}$ \\
$\eta_r(r)\propto r^{-a}$ & $a=1.5$ & & \\
$H(r)$ & $0.6\,r_{\rm subl}\times \left(\frac{r}{r_{\rm subl}}\right)$ & dust & see Sect.~\ref{dust} \\
$r_{\rm o}$ & $70\,r_{\rm subl}$ & & \\ \hline
\end{tabular}
\end{table}

\begin{figure*}
\begin{center}
\includegraphics[angle=0,width=5.5cm]{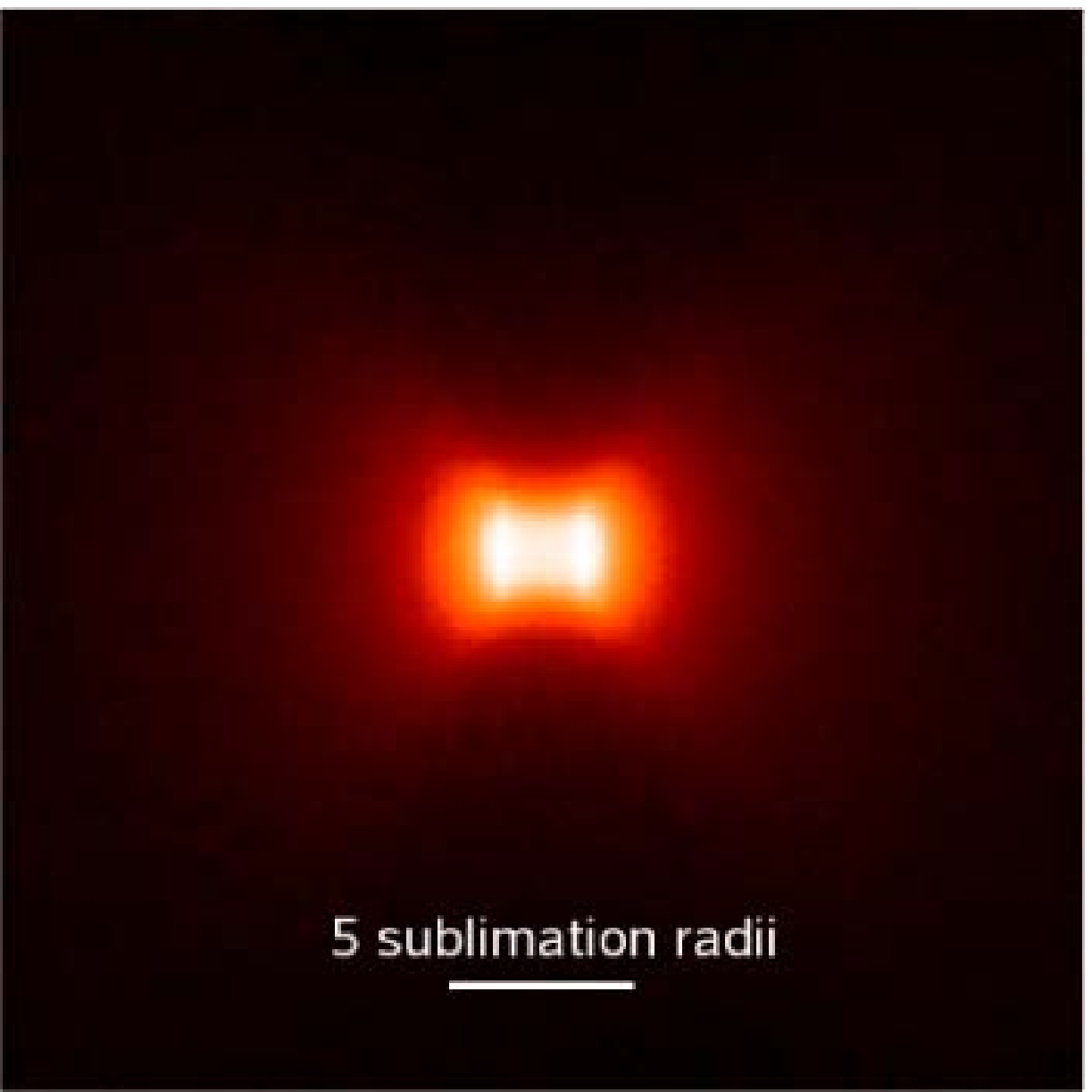}
\includegraphics[angle=0,width=5.5cm]{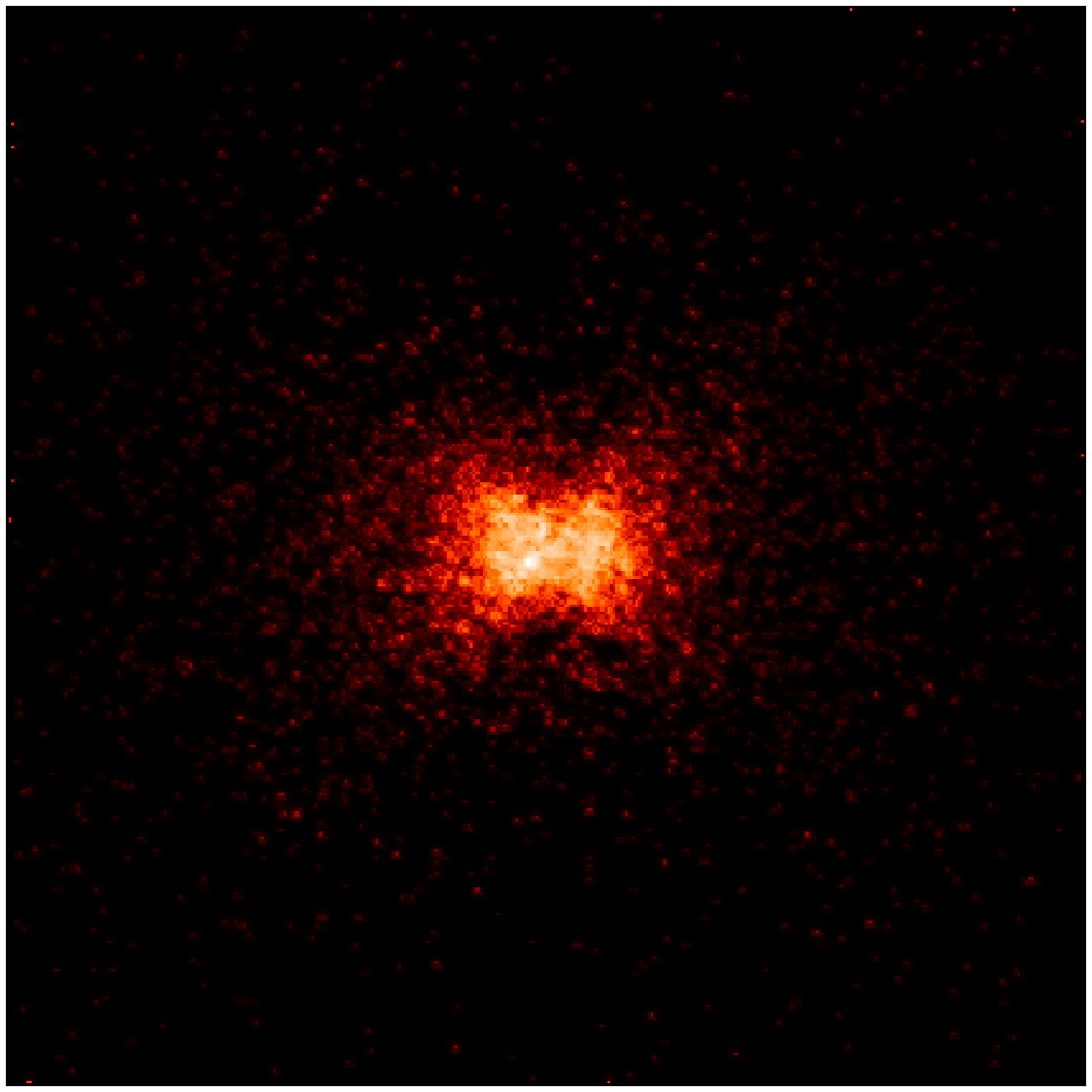}
\includegraphics[angle=0,width=5.5cm]{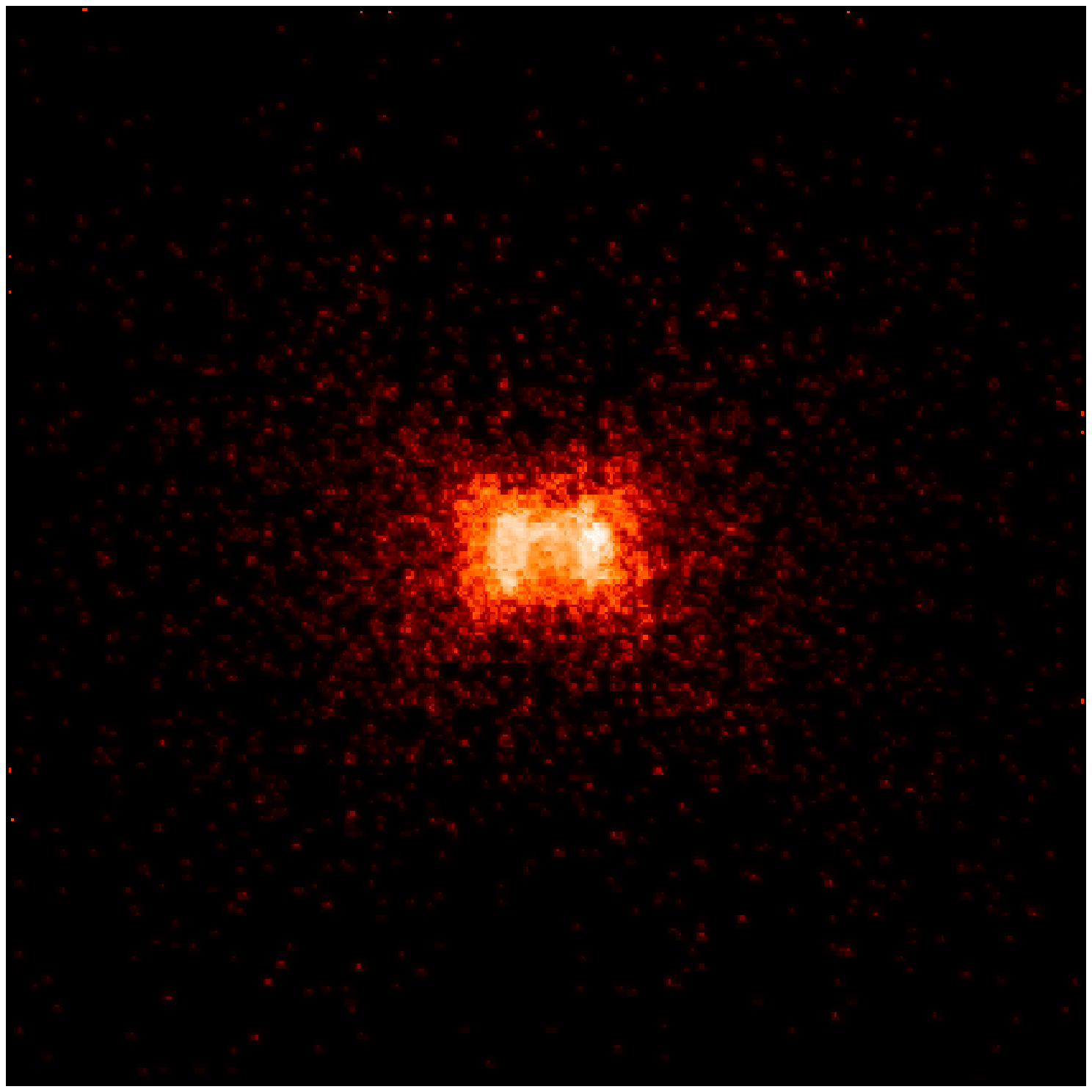}

\includegraphics[angle=0,width=5.5cm]{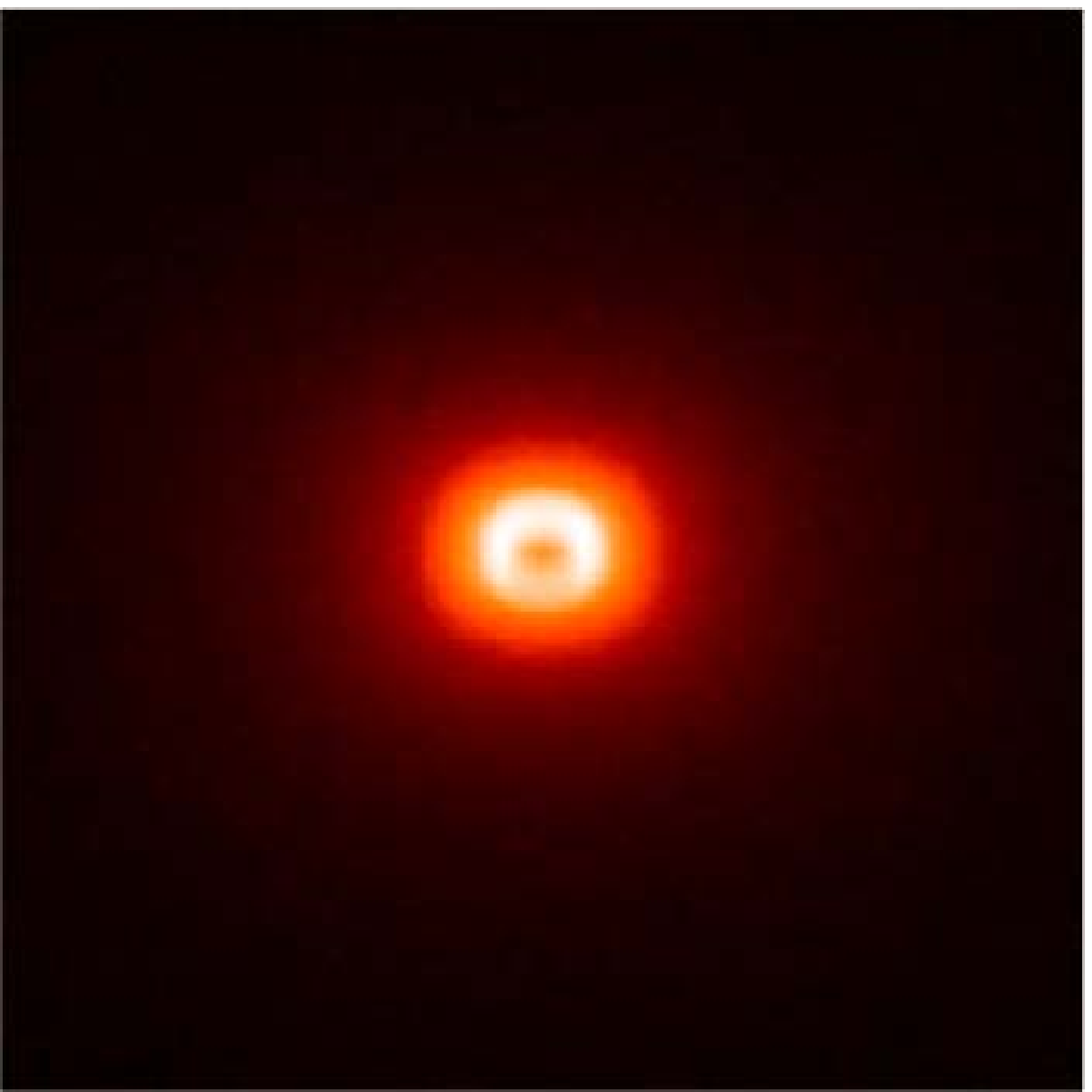}
\includegraphics[angle=0,width=5.5cm]{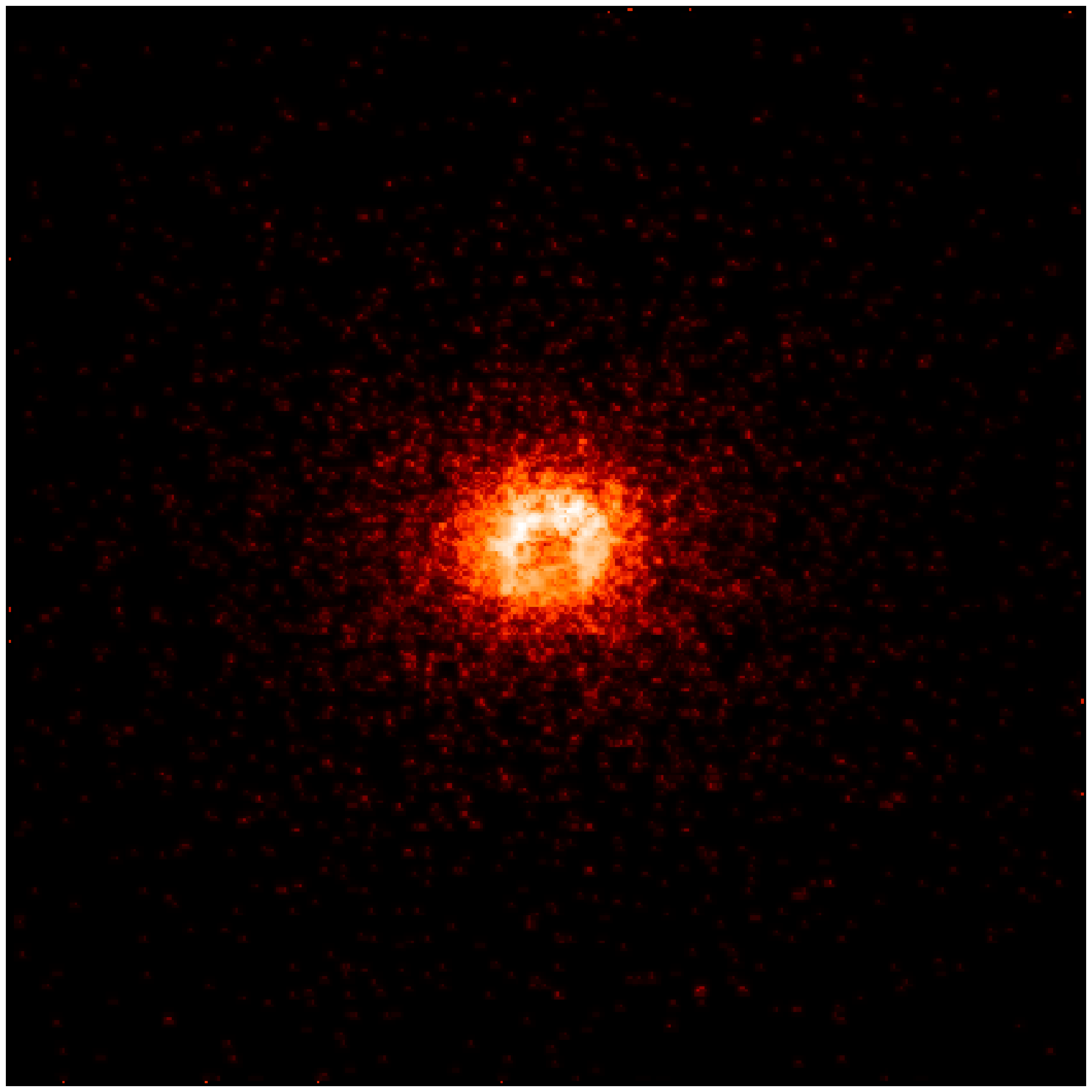}
\includegraphics[angle=0,width=5.5cm]{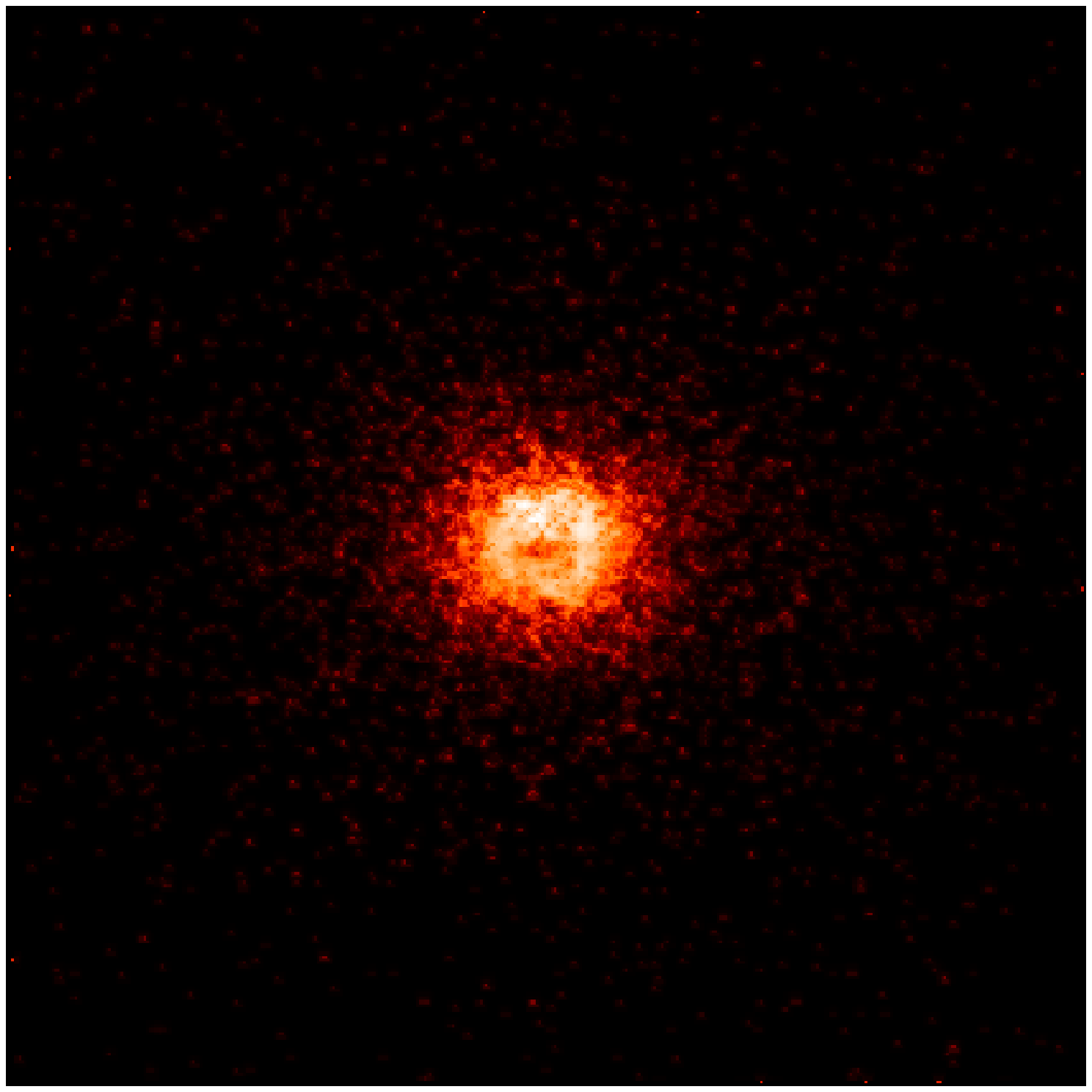}

\includegraphics[angle=0,width=5.5cm]{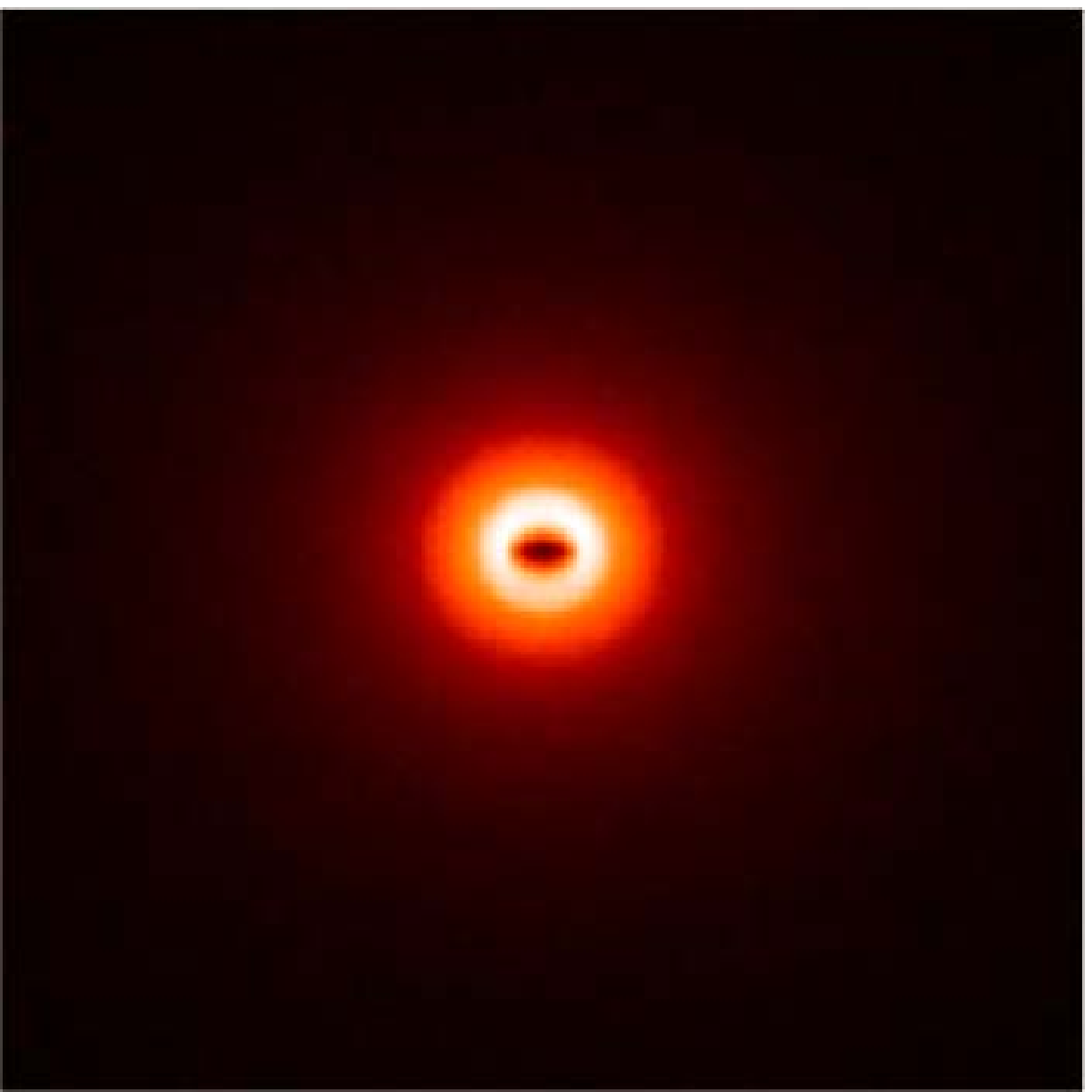}
\includegraphics[angle=0,width=5.5cm]{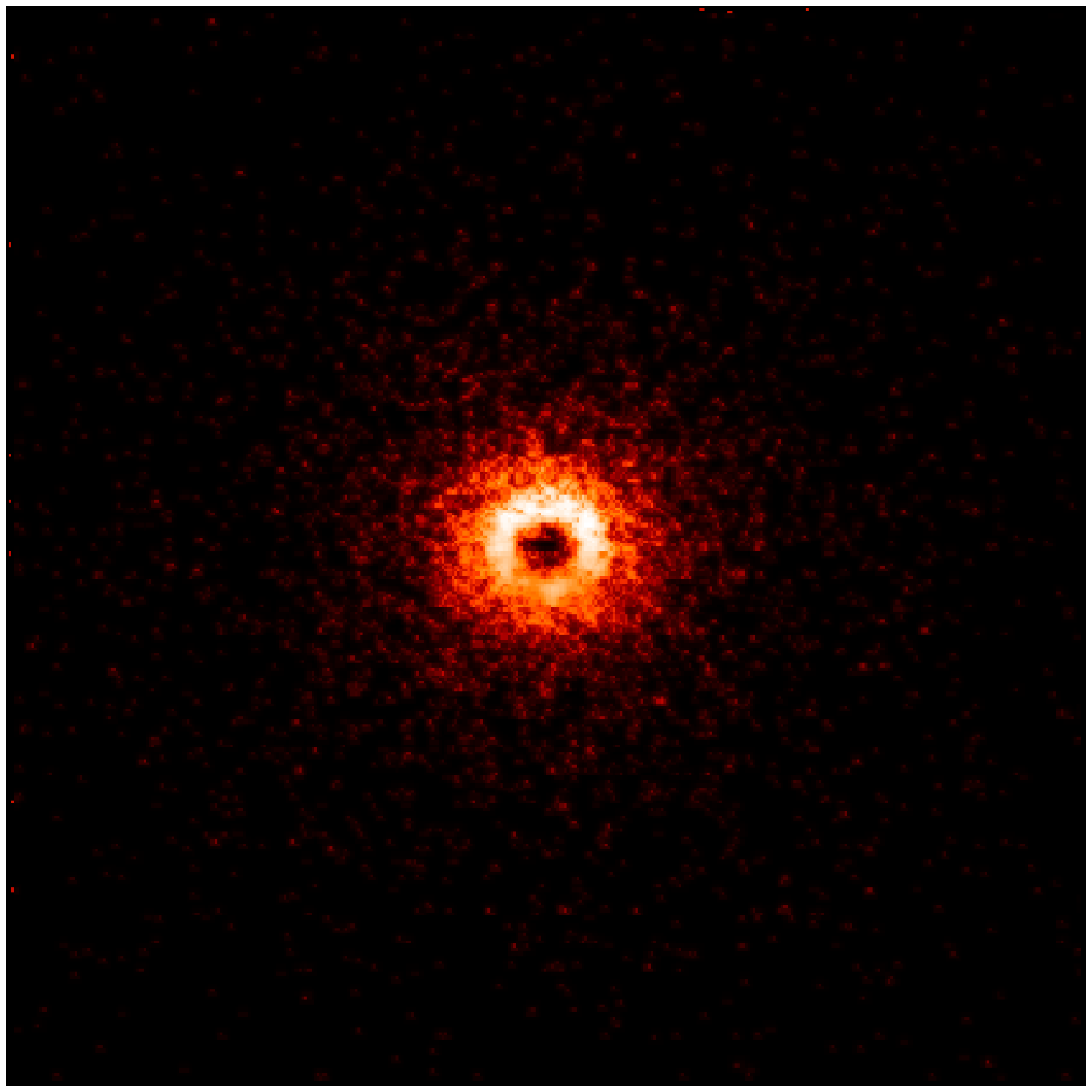}
\includegraphics[angle=0,width=5.5cm]{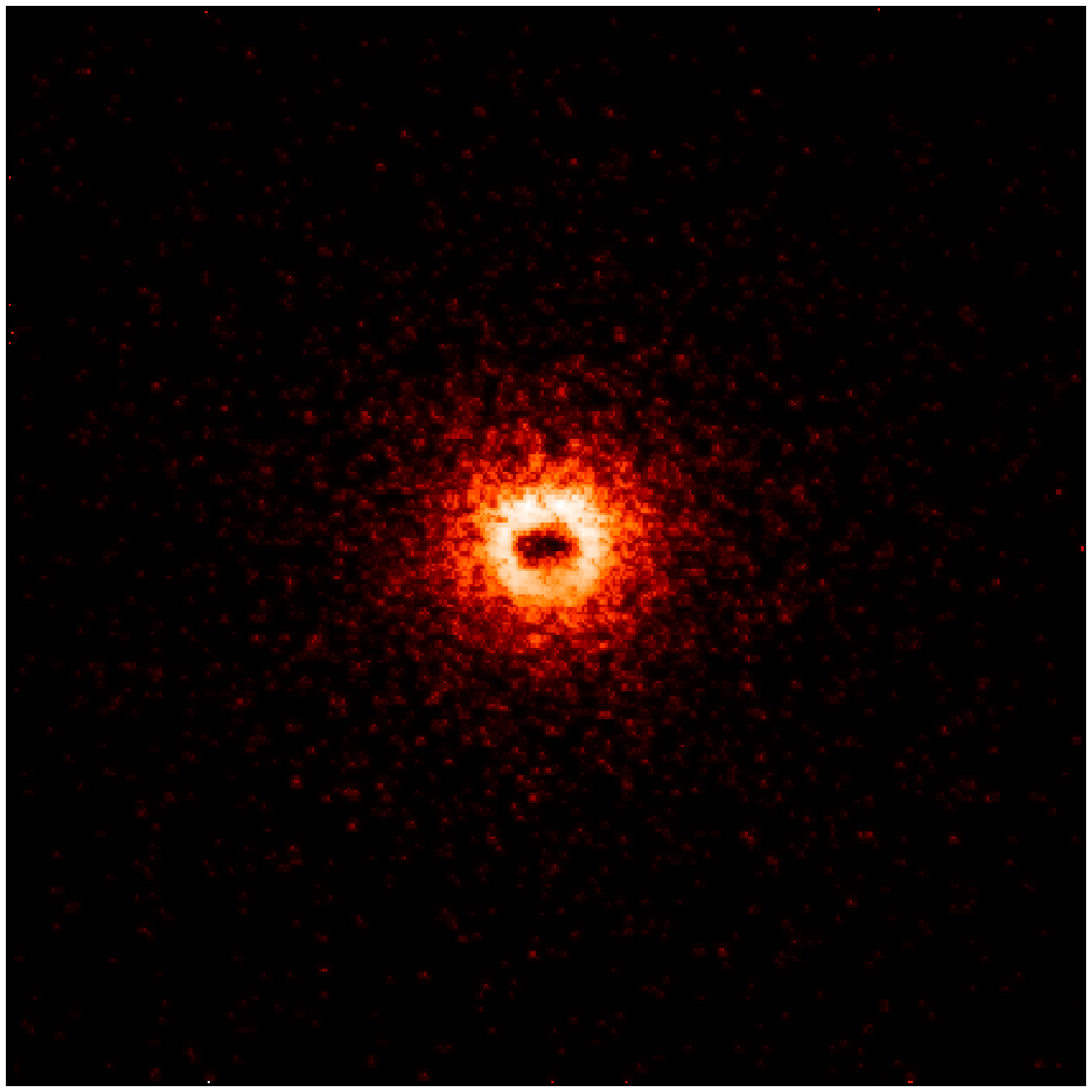}

\includegraphics[angle=0,width=5.5cm]{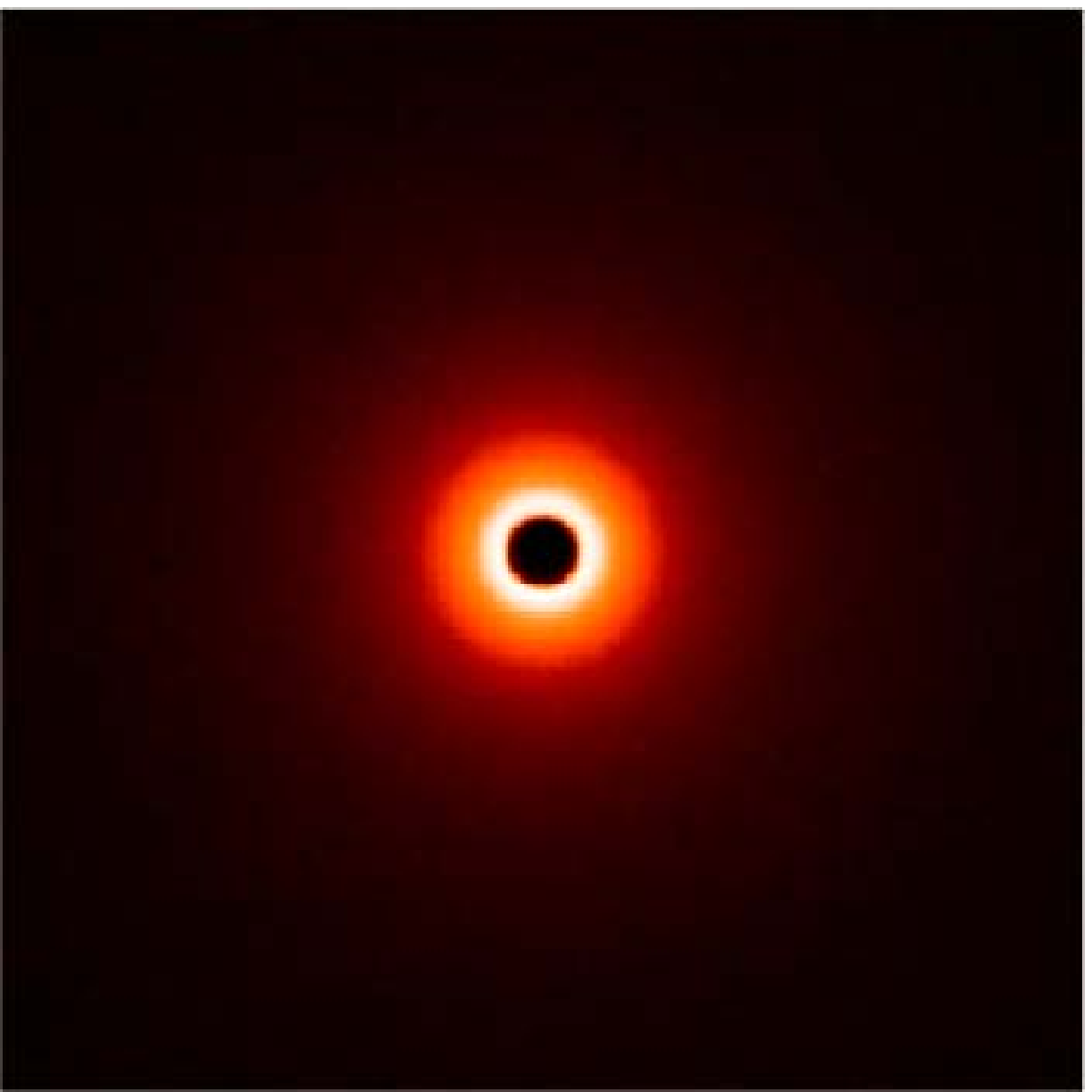}
\includegraphics[angle=0,width=5.5cm]{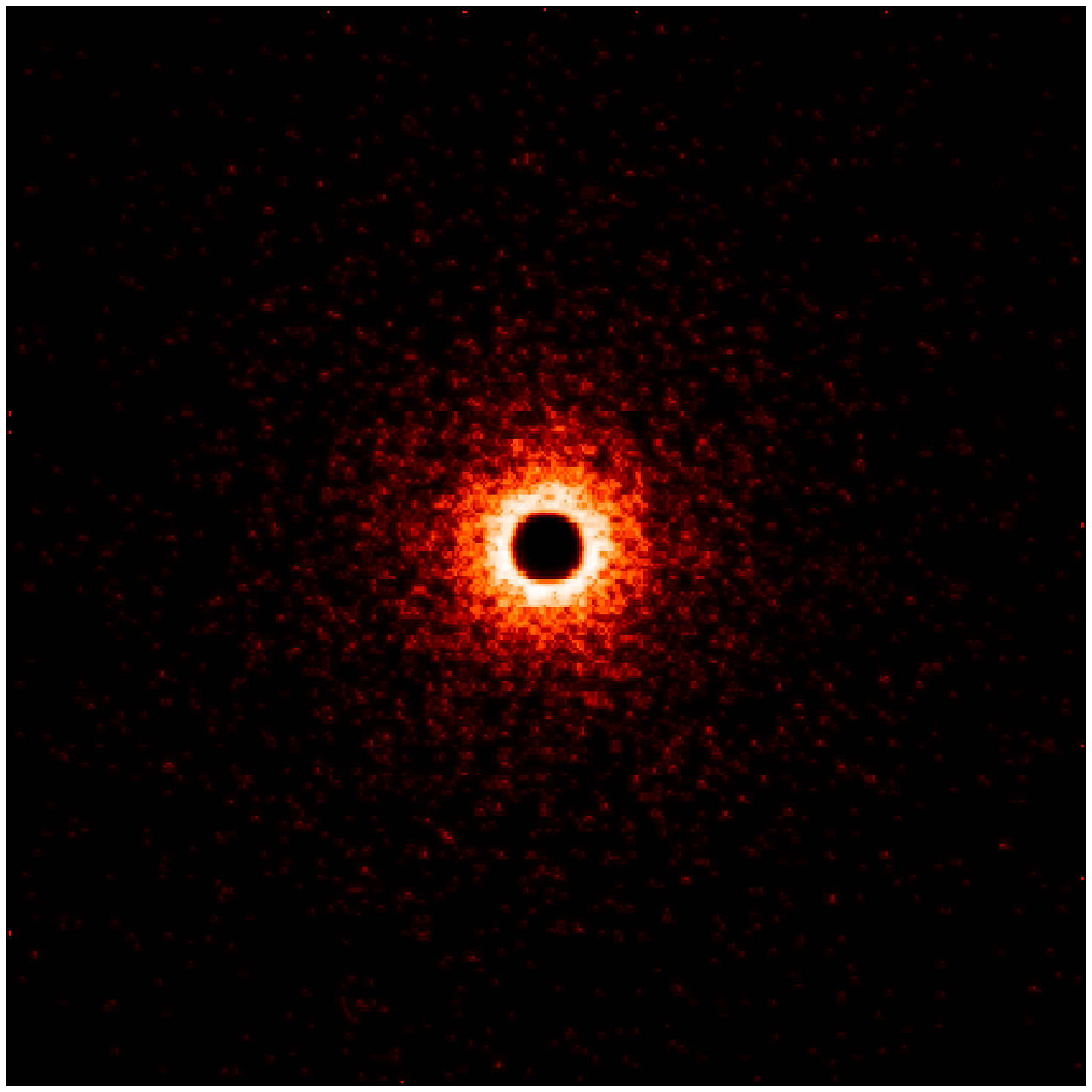}
\includegraphics[angle=0,width=5.5cm]{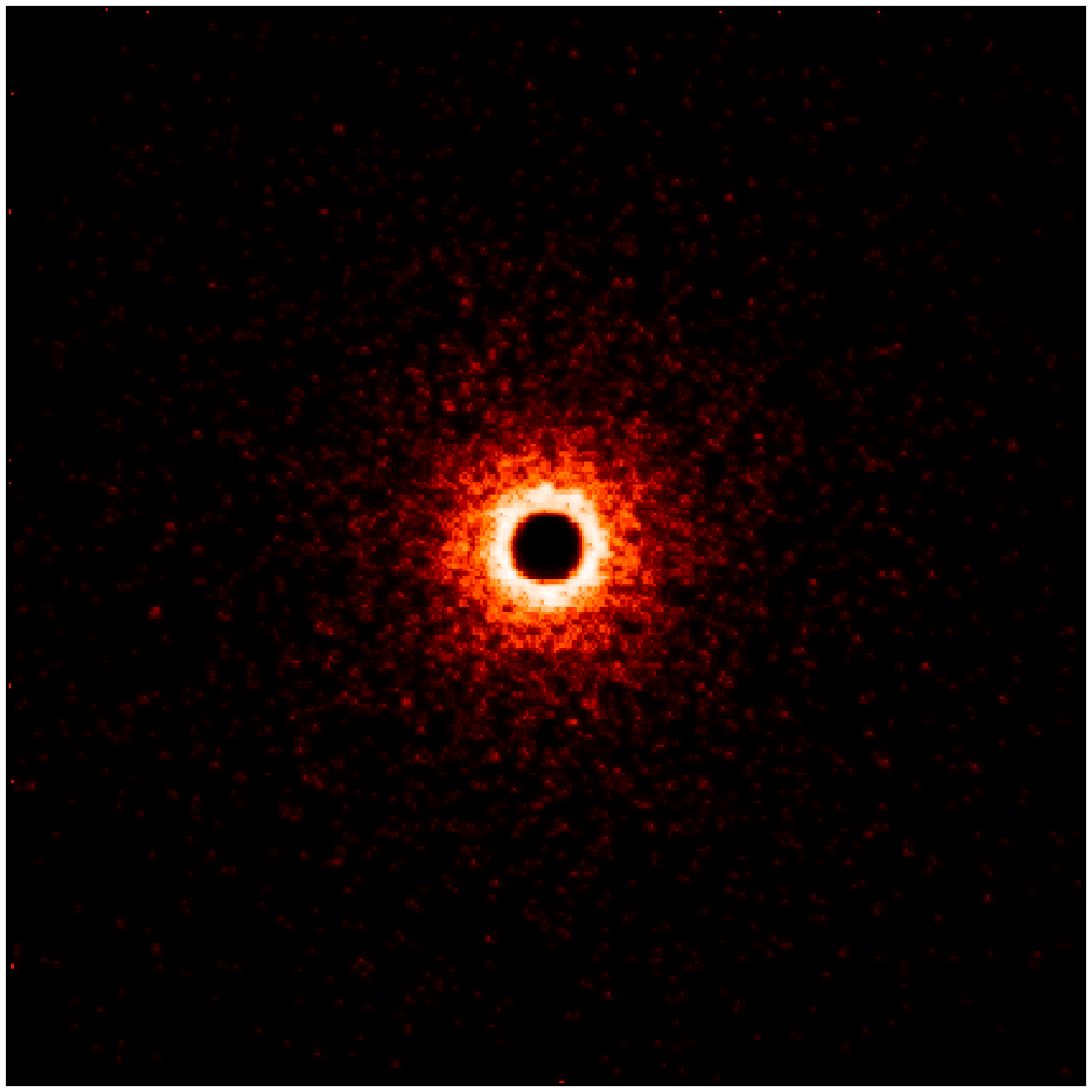}
\end{center}
\caption{$L$-band model images of the torus described by the parameters listed in Table~\ref{tabparstd} (same parameters as used for the SED calculations shown in Fig.~\ref{figVar}). Left column: images obtained by averaging model images of $\sim$200 different random cloud arrangements. Middle and right columns: model images for two particular random cloud arrangements. From top to bottom: $i=90\degr$, $45\degr$, $30\degr$, and $0\degr$.}\label{figImi}
\end{figure*}

\begin{figure}
\sidecaption
\centering
\hspace{0.8cm}
\includegraphics[angle=0,width=5.5cm]{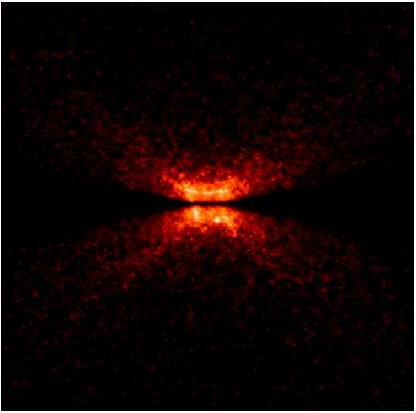}
\caption{\,$K$-band model image for {\it non-isotropic, sinusoidal
radiation} from the accretion disk for one particular random
cloud arrangement, edge-on view, and model parameters listed
in Table~\ref{tabparstd}}\label{Aniso}
\end{figure}

\begin{figure*}
\centering
\includegraphics[angle=0,width=16.5cm]{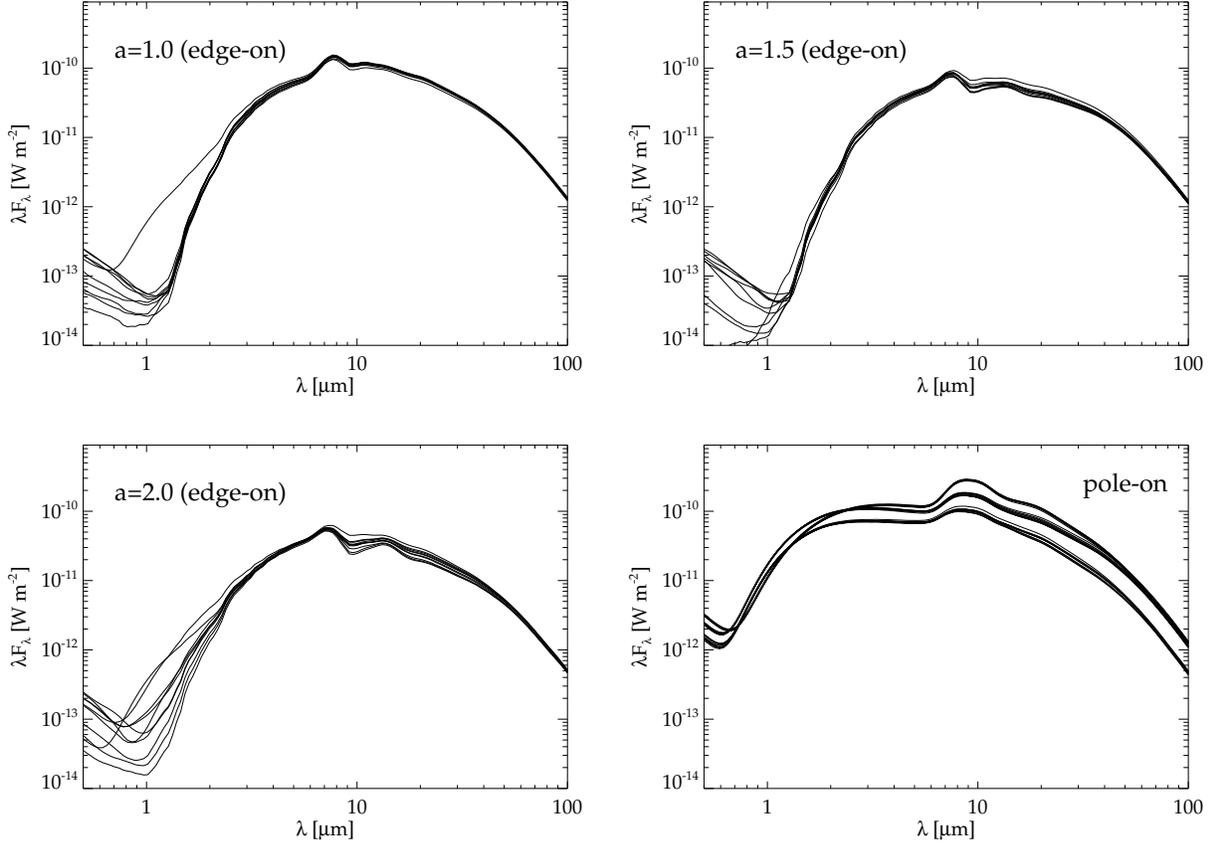}
\caption{The dependence of the SED on $\eta_r$ (model parameters in Table \ref{tabparstd}). The upper panels show edge-on SEDs for 10 different random cloud arrangements for $a=1.0$ (left) and $a=1.5$ (right). In the lower left panel we present the edge-on view for $a=2.0$, while the lower right panel combines pole-on views of the three models with $a=1.0$ (top), $a=1.5$ (middle), and $a=2.0$ (bottom).}\label{figCD}
\end{figure*}

\subsection{Statistical cloud distribution}\label{statcd}
\begin{figure*}
\centering
\includegraphics[angle=0,width=16.5cm]{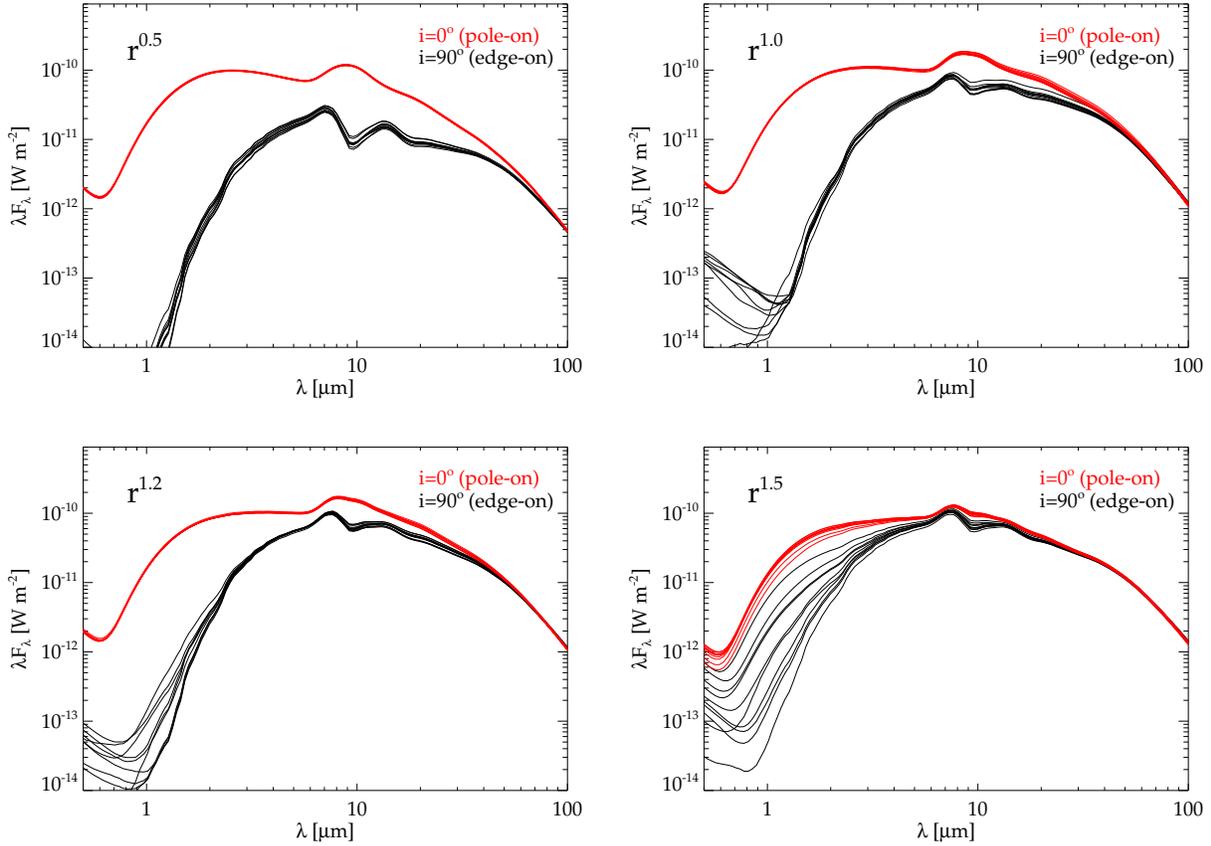}
\caption{SEDs for different flaring (see text; model parameters in Table \ref{tabparstd}). From upper left to lower right: $H \propto r^{0.5}$, $r^{1.0}$, $r^{1.2}$, and $r^{1.5}$, respectively. Each panel consists of SEDs for 10 different random cloud arrangements seen edge-on (black) and pole-on (red).}\label{figSH}
\end{figure*}

The distribution of clouds follows a statistical process. Thus, it is expected that different statistical realizations of this distribution will change the torus SED. Since recent studies of clumpy tori use statistical averages \citep{Nen02,Nen05}, it is also worth investigating the scatter that has to be expected. In Fig. \ref{figVar} we show the SED variation for one set of parameters (listed in Table \ref{tabparstd}). The only difference between the SEDs is the random arrangement of the clouds, following the same distribution function $\eta(r,z,\varphi)$. Analyzing the data shows that the slope of the NIR flux is comparable for all random cloud arrangements. The absolute fluxes and the silicate feature differ by a factor of $2-3$ for the edge-on view. The large SED scatter at $i=30\degr$ is a result of the transition from edge-on to pole-on view. The FIR fluxes get close to each other at longer wavelengths since the optical depth of the torus becomes smaller than unity, so that obscuration no longer plays an important role.

We investigated the origin of the dependence of the SED on the random cloud arrangement. As we found out, these variations are caused by the FOCs. They are the main contributors at the shorter wavelengths of the SED. A few bright FOCs at high latitudes can change the SED considerably since high-latitude clouds are usually less obscured.

In Fig. \ref{figImi} we present model images for inclinations $i=90\degr$, $45\degr$, $30\degr$, and $0\degr$. For each inclination angle, we show two different kinds of images: the left column presents an average image derived from $\sim$200 different random cloud arrangements; the mid and right columns show images calculated for two individual random cloud arrangements (wavelength range $4.65-4.95\,{\rm\mu m}$; for model parameters as listed in Table~\ref{tabparstd}). In all images, the assumed faint emission from the central AGN is not shown (see Sect. \ref{AGNrad} and Fig. \ref{fig1}). The images obtained for particular random cloud arrangements show a similar overall appearance with distinct differences in the substructure. The {\it grainy structure} in the images is a result of the {\it clumpiness} of the torus and not related to noise of the radiative transfer simulations. It illustrates the appearance of a clumpy torus for very high-angular resolution observations.

In the average image for $i=90\degr$, a polar cavity with an X-shaped structure is visible. At $i=45\degr$ and $90\degr$, a bright inner structure is the dominant source of emission, which is caused by the inner torus rim. At $i=30\degr$, parts of the inner rim of the torus can be seen as a crescent structure. The central dark region for $i=30\degr$ and $0\degr$ is the inner dust-free region. The apparently well-defined brightness steps in the averaged images will disappear as more different clouds are added to the database of SOCs.

The model images presented in Fig.~\ref{figImi} were calculated for an isotropic radiation of the accretion disk. Fig.~\ref{Aniso} shows a model image obtained for a {\it non-isotropic}, sinusoidal radiation field of the accretion disk. If the torus is illuminated by this non-isotropic radiation, a dark dust lane appears in the disk plane when the torus is viewed edge-on.

\subsection{Radial Distribution function}\label{radfunc}

To illustrate the effect of the radial cloud distribution function $\eta_r(r)$ on the SED, we keep all parameters in Table~\ref{tabparstd} constant, except for exponent $a$ of the radial power law. Fig. \ref{figCD} shows the resulting SEDs for $a=1.0$, $a=1.5$, and $a=2.0$ for both the edge-on and pole-on views of the torus. The relative depth of the silicate feature in absorption becomes larger with growing $a$. Additionally, the NIR flux decreases for both the pole-on and edge-on views, and the variations become stronger. This can be explained by the higher density of clouds with large $\tau_{cl}$ close to $r_{\rm subl}$. As a result, the obscuration of the FOCs in that area becomes higher and their total flux decreases. Small differences in the arrangement may then lead to higher variations.

It is interesting to look at the evolution of the silicate emission feature in the pole-on view. The feature is quite flat and less pronounced than for smooth dust distributions, and it decreases for larger $a$. This effect comes from the higher cloud density of FOCs close to $r_{\rm subl}$ when emission and absorption in that region become equally important. Tori with $a\ga 2$ can flatten out the emission feature for small $i$ or even show moderate absorption.

\subsection{The effect of flaring}\label{flare}

The $z$-dependence of the cloud distribution function is a Gaussian profile $\eta_z \propto \exp\left(-z^2/2H(r)^2\right)$ (see Sect.~\ref{TorModels}). Thus, the toroidal structure flares depending on the $r$-dependence of $H$. In Fig. \ref{figSH} we show the model SEDs for $H \propto r^{0.5}$, $r^{1.0}$, $r^{1.2}$, and $r^{1.5}$. For the edge-on geometry, the flaring has some effect on the relative depth of the silicate feature. With small flaring the feature is deeper, and the overall flux is also lower by a factor of $\sim$$3-4$. This can be explained by the absence of clouds at higher latitudes which are usually less obscured and thus contribute significantly to the NIR and MIR fluxes.
\begin{figure}
\centering
\includegraphics[angle=0,width=8.5cm]{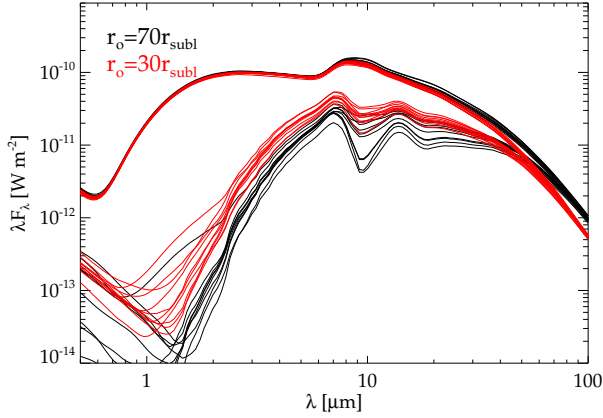}
\caption{SEDs for outer radii of $r_{\rm o}=70\, r_{\rm subl}$ (black) and $r_{\rm o}=30\, r_{\rm subl}$ (red) seen edge-on and pole-on (model parameters in Table \ref{tabparstd}).}\label{figRo}
\end{figure}

For the pole-on view, the importance of flaring is more evident. The NIR and MIR fluxes decrease significantly when the flaring becomes stronger since there are more and more clouds at high latitudes that obscure the denser parts of the torus structure. As a result, the emission feature vanishes for $H \propto r^{1.5}$ and can even turn into weak absorption for stronger flaring.

\subsection{The outer radius of the torus}

Since the main torus emission in the NIR and MIR comes from the innermost part, it is expected that the SED is less sensitive to the outer boundary of the torus. This especially applies if $\tau_{cl}$ decreases with $r$, as proposed by the accretion scenario (Sect.~\ref{AccScen}). In this case, the innermost part of the torus dominates since absorption by more distant clouds is weak. Thus, the strongest effect of the outer torus radius $r_{\rm o}$ is expected in cases where distant clouds significantly contribute to the obscuration of the inner torus. To test this, we investigated the influence of $r_{\rm o}$ for the extreme case of $\tau_{cl}=70{\rm\,(constant)}$ throughout the torus. We simulated SEDs for $r_{\rm o}=30\, r_{\rm subl}$ and $r_{\rm o}=70\, r_{\rm subl}$, respectively.

Our results are shown in Fig. \ref{figRo}. The overall shape of the SED is similar for both model $r_{\rm o}$, while there are some differences in the absolute fluxes. Since the central region is less obscured for $r_{\rm o}=30\, r_{\rm subl}$, the NIR and MIR fluxes are higher. On the other hand, the FIR flux of the $r_{\rm o}=70\, r_{\rm subl}$ model is above the flux for the smaller radius since the outer cold part is missing in the latter model.

\section{Simultaneous modeling of the SED and visibilities of NGC\,1068}\label{ModSEDN1068}

The goal of our radiative transfer modeling studies of NGC\,1068 is to find a clumpy torus model which can simultaneously reproduce both (1a,b) the observed high spatial resolution IR SED and (2a--c) the interferometric visibilities obtained during the last few years:

(1a) SED values in the wavelength range of 1.3 to 18 $\mu$m listed in Table \ref{tabobs1068},

(1b) high angular resolution MID IR SED in the wavelength range from 8 to 13 $\mu$m obtained using the VLTI MIDI instrument \citep{Jaf04},

(2a) speckle interferometric $H$- and $K$-band visibilities \citep{Wei04},

(2b) MIR visibilities ($8-13\,\rm{\mu m}$) obtained with the VLTI MIDI instrument \citep{Jaf04},

(2c) $K$-band visibilities obtained with the VLTI VINCI instrument \citep{Wit04}.

\begin{figure}
\centering
\includegraphics[angle=90,width=9.0cm]{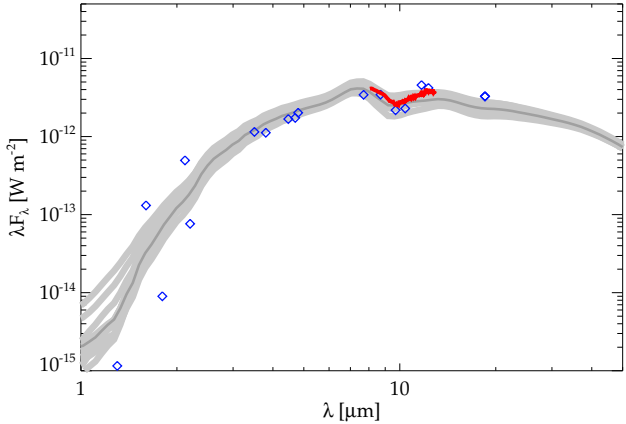}
\caption{Comparison between the observed high-resolution SED of NGC\,1068 (blue diamonds and red curve) and our model SED (model parameters in Table 5). The red line represents the total MIDI flux \citep{Jaf04}. The shaded area shows the range of model SED variations obtained for 10 different random cloud arrangements. The dark grey line is the average of the 10 individual model SEDs.}\label{figN1068}
\end{figure}

We tried to find a set of model parameters that can simultaneously reproduce all of these observations. For this goal we scanned the parameter space, as described in Sect.~\ref{Results}, and obtained the clumpy torus model parameters listed in Table~\ref{tabpar1068}. Figs.~\ref{figN1068} and \ref{figImN1068} present the model SED and $K$-, $M$-, and $N$-band model images.
\begin{table}
\caption{Photometric data for the nuclear region of NGC\,1068 from high-resolution observations.}\label{tabobs1068}
\centering
\vspace{0.1cm}
\begin{tabular}{c c c c}
\hline\hline
Wavelength & Flux & Phot. Aperture & Reference \\
$\rm\mu m$ & Jy & arcsec & \\ \hline
1.3 & $0.00048$ & 0.2 & R98 \\
1.6 & $0.07 \pm 0.02$ & 0.1 & W04 \\
1.8 & $0.0054$ & 0.2 & R98 \\
2.1 & $0.35 \pm 0.09$ & 0.1 & W04 \\
2.2 & $0.056$ & 0.27 & P04 \\
3.5 & $1.34$ & 0.4 & M00 \\
3.8 & $1.42$ & 0.4 & M03 \\
4.5 & $2.5$ & 0.27 & P04 \\
4.7 & $2.72 \pm 0.14$ & 0.4 & M03 \\
4.8 & $3.23 \pm 0.32$ & 0.4 & M00 \\
7.7 & $8.8 \pm 1.0$ & 0.4 & T01 \\
8.7 & $10.0 \pm 1.2$ & 0.4 & T01 \\
9.7 & $7.04 \pm 0.80$ & 0.4 & T01 \\
10.4 & $7.94 \pm 0.87$ & 0.4 & T01 \\
11.7 & $17.8 \pm 2.2$ & 0.4 & T01 \\
12.3 & $17.2 \pm 2.1$ & 0.4 & T01 \\
18.5 & $20.2 \pm 3.4$ & 0.4 & T01 \\ \hline
\multicolumn{4}{l}{R98 \citep{Rou98}, W04 \citep{Wei04},}\\
\multicolumn{4}{l}{P04 \citep[Prieto, published in][]{Sch05},}\\
\multicolumn{4}{l}{M00 \citep{Mar00}, M03 \citep{Mar03},} \\
\multicolumn{4}{l}{T01 \citep{Tom01}} \\
\end{tabular}
\end{table}

\begin{table}
\caption{Parameters of our NGC\,1068 clumpy torus model.}\label{tabpar1068}
\centering
\vspace{0.1cm}
\begin{tabular}{l l | l l}
\hline\hline
Torus & Model Value & Clouds & Model Value \\ \hline
$L_{\rm AGN}$ & $0.24\times\,L_{12}$ & $R_{cl}(r)$ & $0.01\,{\rm pc}\times \left(\frac{r}{r_{\rm subl}}\right)^{1.5}$ \\
$r_{\rm subl}$ & $0.28\,{\rm pc}$ & $N_0$ & 10 \\
$d$ & $14.4\,{\rm Mpc}$ & $\tau_{cl}$ & $40\times\left(\frac{r}{10\,r_{\rm subl}}\right)^{-0.8}$ \\
$\eta_r(r)\propto r^{-a}$ & $a=1.5$ & & \\
$H(r)$ & $0.6\,r_{\rm subl}\times \left(\frac{r}{r_{\rm subl}}\right)$ & dust & see Sect.~\ref{dust} \\
$r_{\rm o}$ & $70\,r_{\rm subl}$ & & \\
$i$ & $55\degr$ & & \\ \hline
\end{tabular}
\end{table}

Fig.~\ref{figN1068} shows a comparison of NGC\,1068's average model SED, the variation range of the model SEDs obtained for 10 different random cloud arrangements, and the observed high angular resolution SED. There is a satisfactory agreement between the observed and modeled SED.

Fig.~\ref{figN1068vis} shows the visibilities of our NGC\,1068 model (Table \ref{tabpar1068}) in the range of $8-13\,{\rm \mu m}$. The model visibilities are roughly in agreement with the observed MIDI visibilities for 42~m and 78~m baselines \citep{Jaf04}. The discrepancies are possibly caused by the difference between our assumed dust properties (see Sect.~\ref{dust}) and the real ones.

\begin{figure}
\centering\includegraphics[angle=90,width=8.7cm]{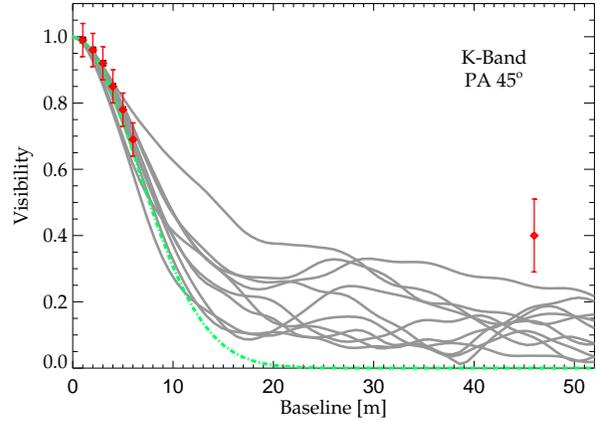}
\caption{Comparison between the $K$-band model visibilities (grey lines; model parameters of Table~\ref{tabpar1068}) for 10 different random cloud arrangements and the observed visibilities (red symbols) at 0-6~m and 46~m baseline at PA 45\degr. The green dashed-dotted line shows a Gaussian profile with 28~mas FWHM (=2.0 pc; see Table \ref{N1068size}).}\label{visk}
\end{figure}

\begin{figure*}
\includegraphics[angle=90,width=13.5cm]{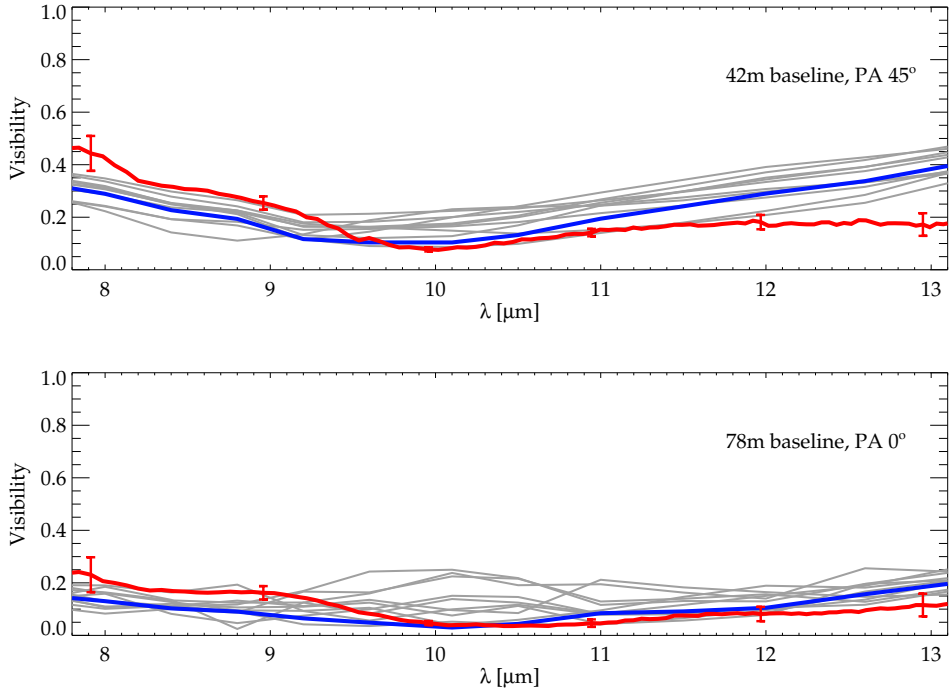}
\caption{\,Comparison between the observed NGC\,1068 MIDI visibilities (red line) at projected baselines of 42~m (upper panel) and 78~m (lower panel), and our model visibilities (model parameters in Table~\ref{tabpar1068}). The grey lines show the visibilities for 10 different random cloud arrangements. The blue line represents the average model visibility.} \label{figN1068vis}
\end{figure*}

\begin{figure*}
\begin{center}
\includegraphics[angle=0,width=5.5cm]{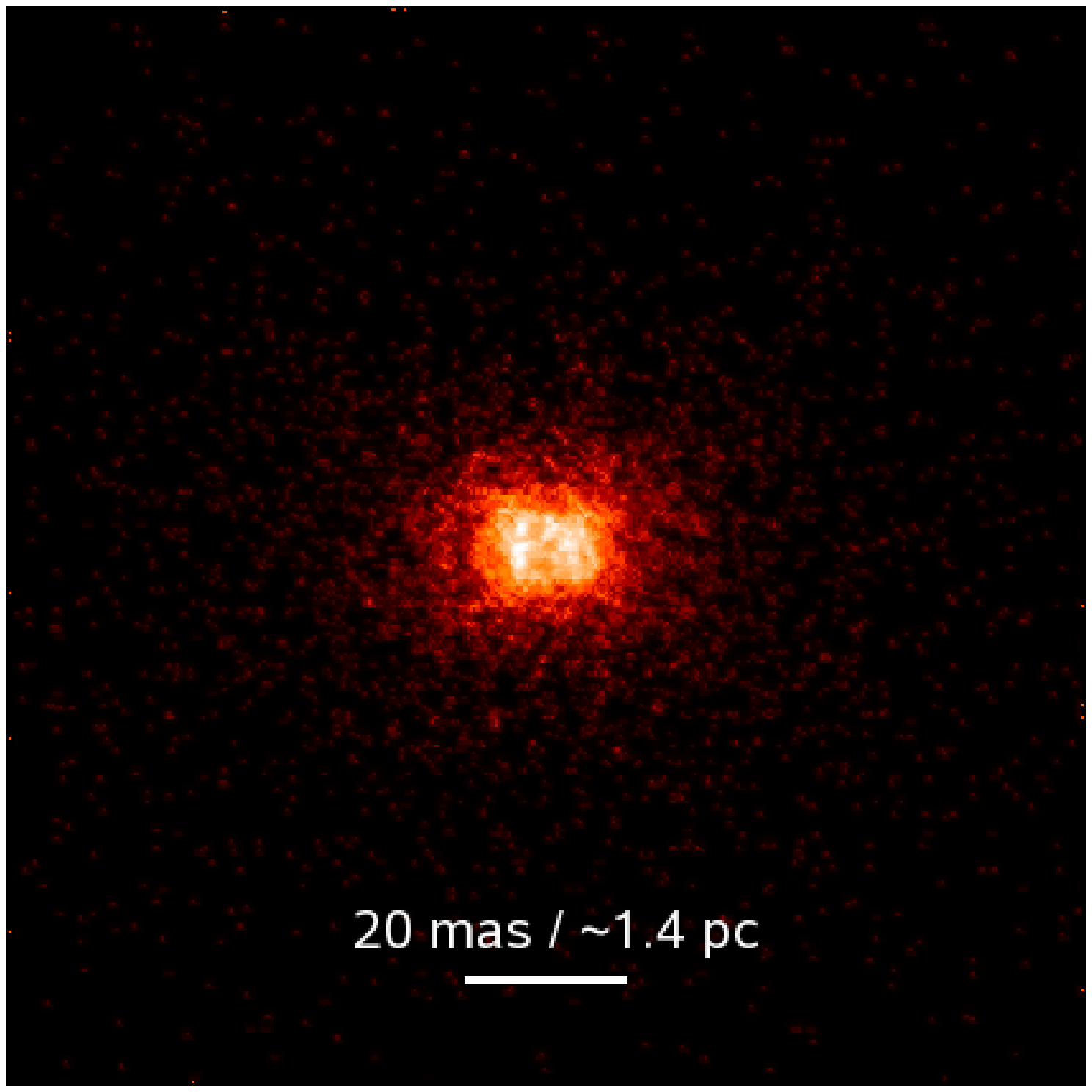}
\includegraphics[angle=0,width=5.5cm]{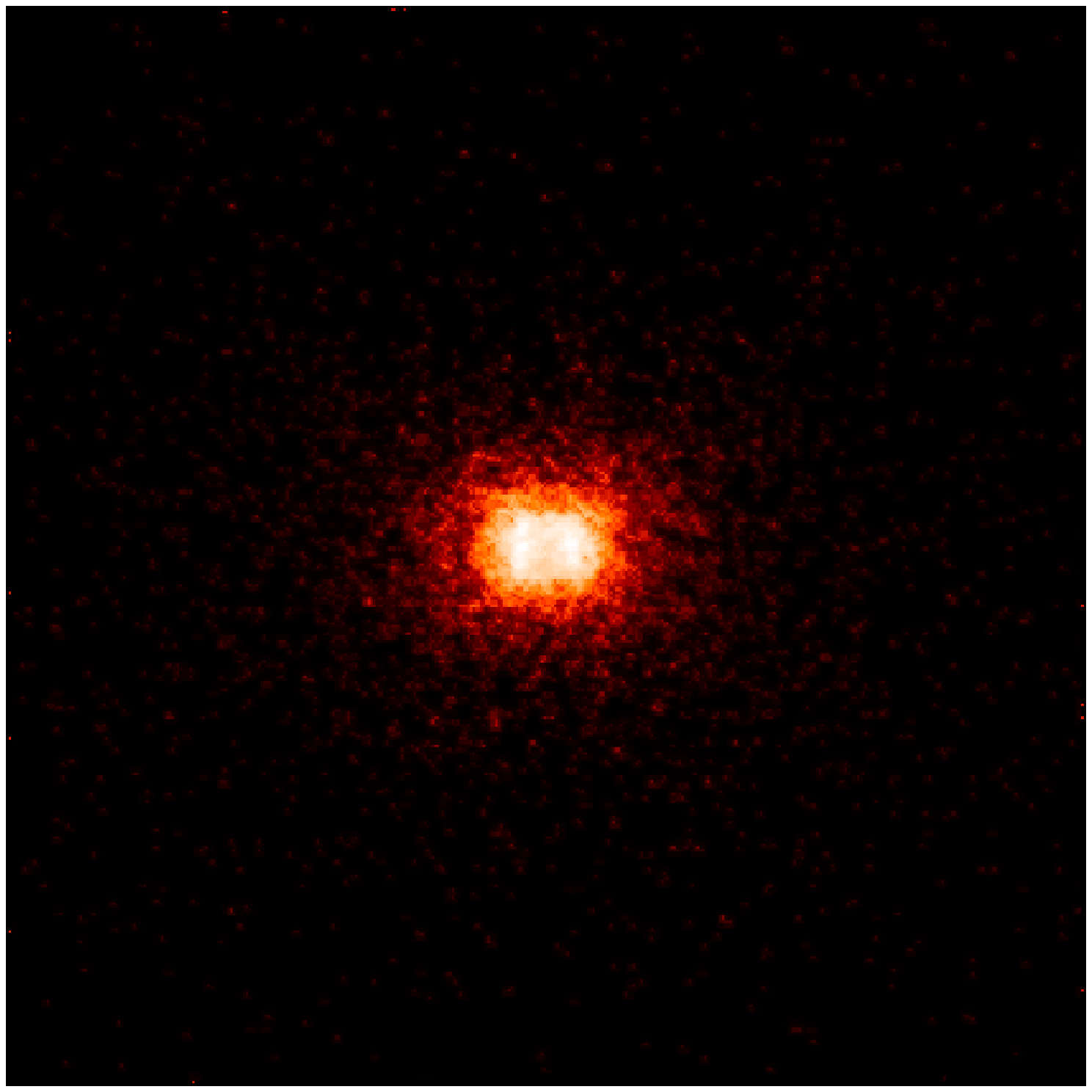}
\includegraphics[angle=0,width=5.5cm]{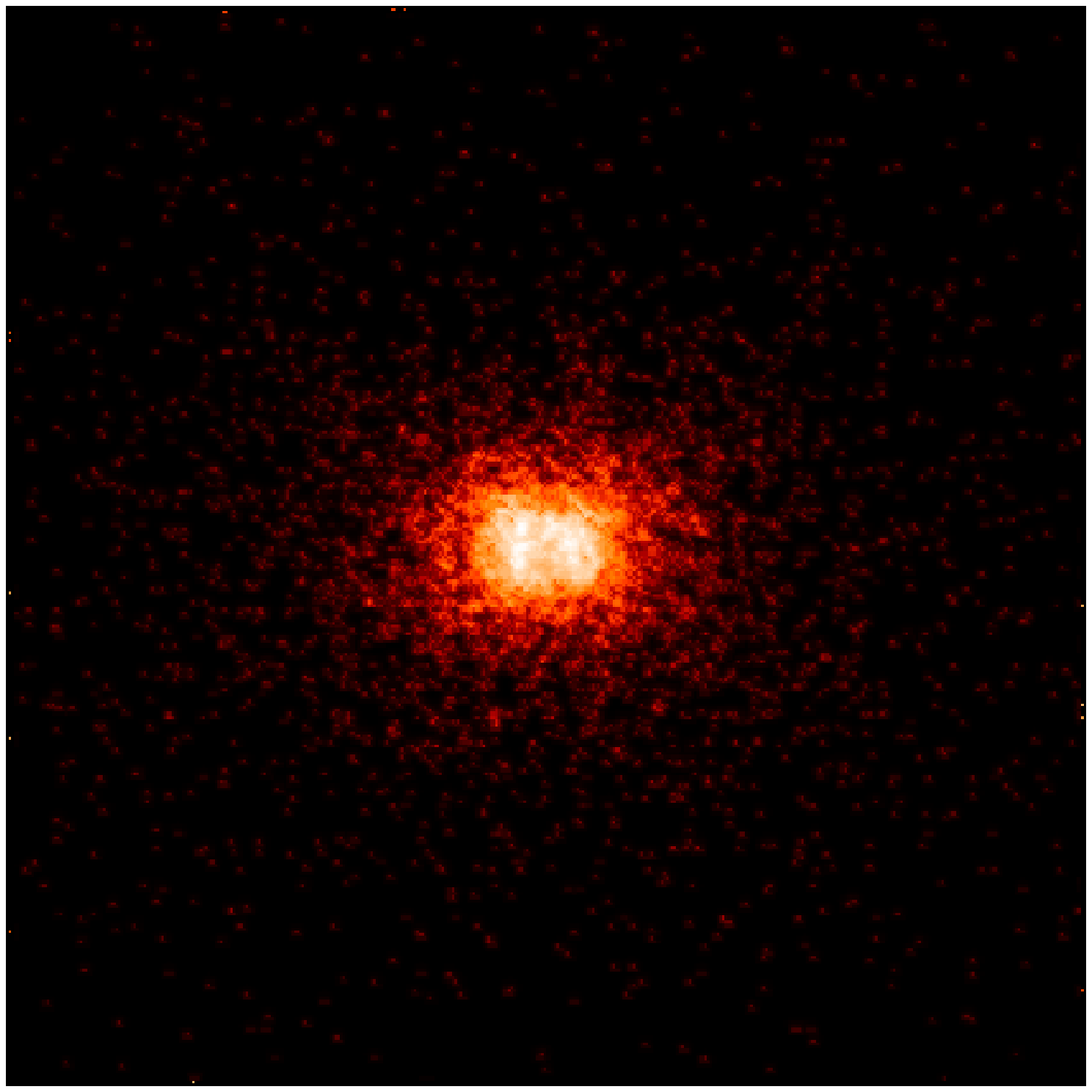}

\includegraphics[angle=0,width=5.5cm]{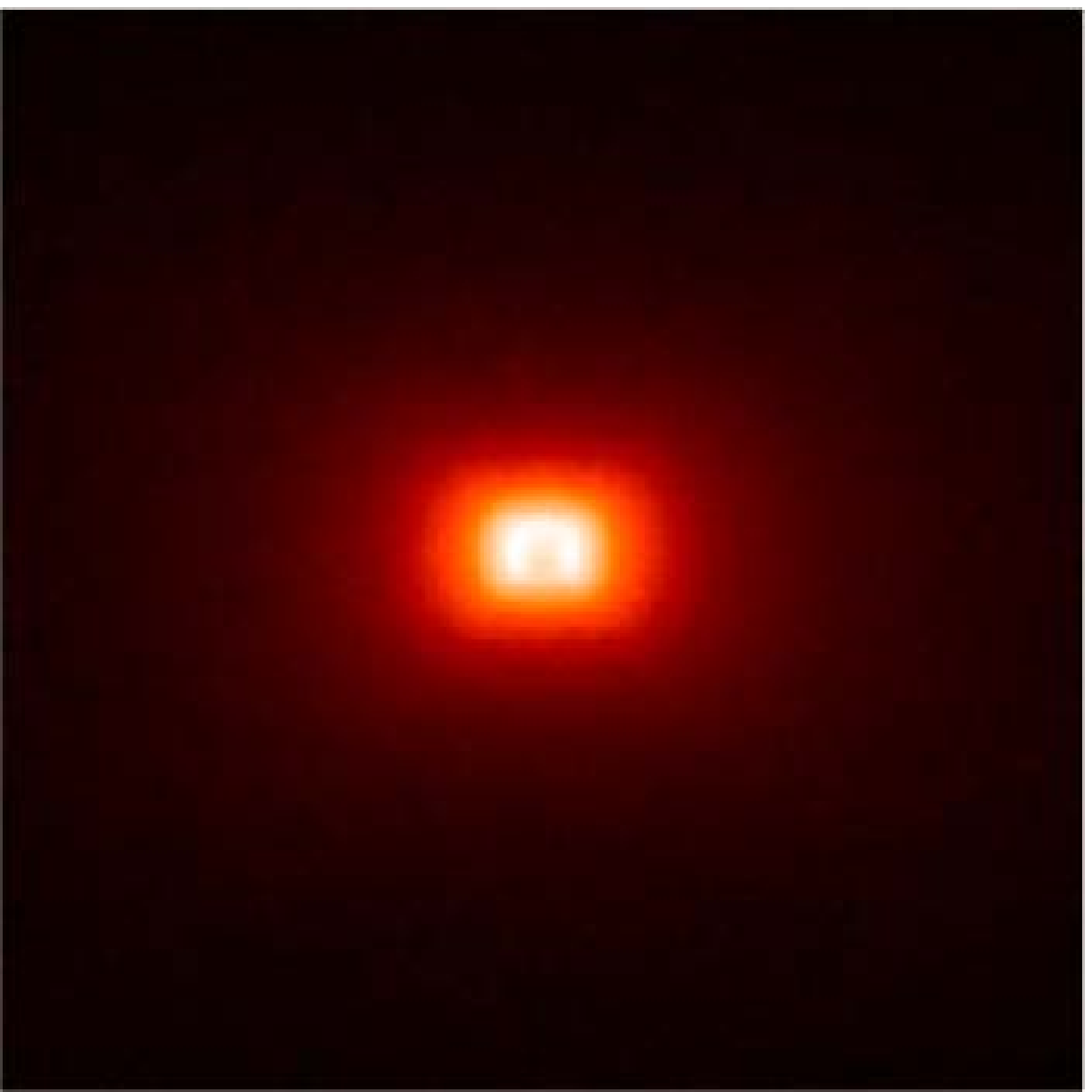}
\includegraphics[angle=0,width=5.5cm]{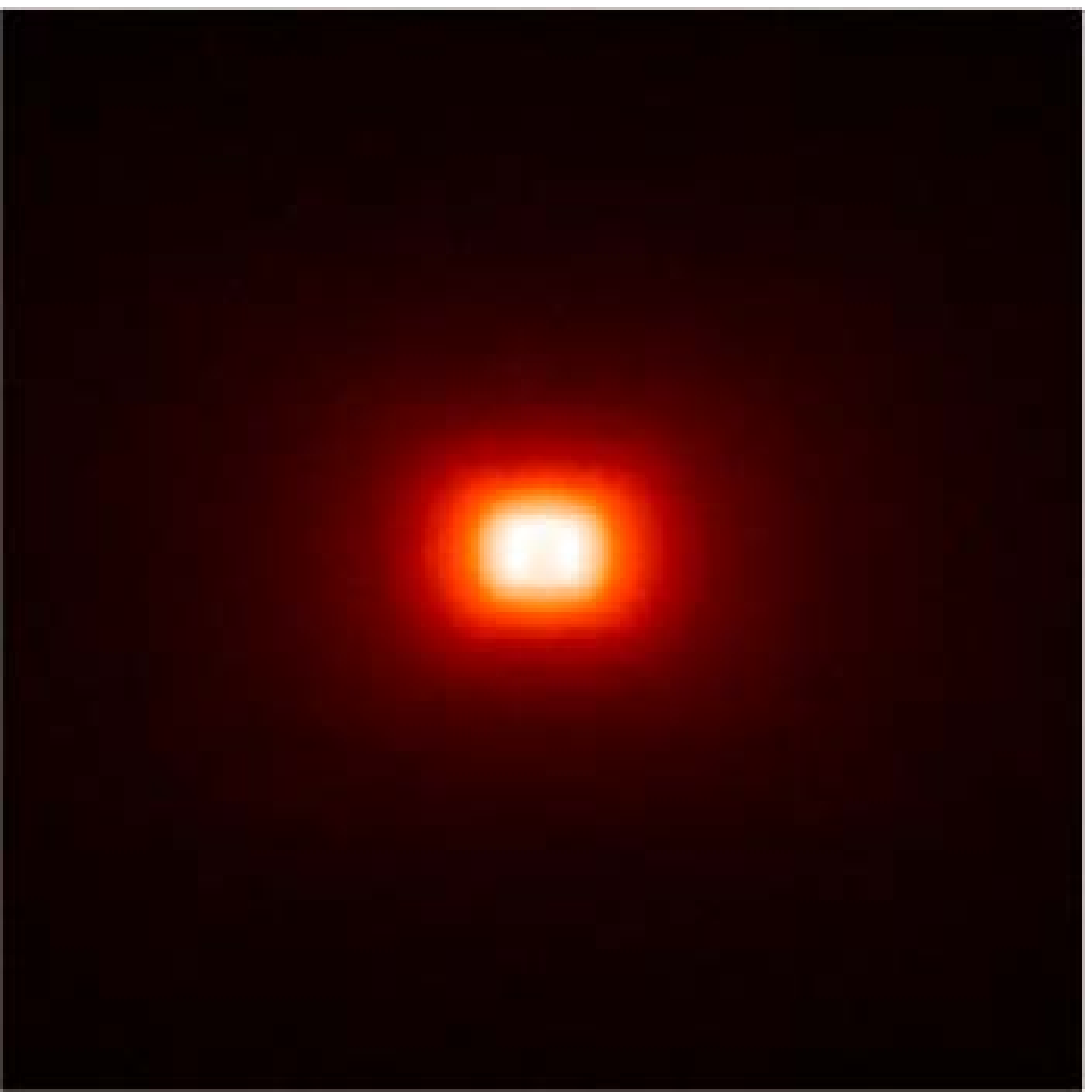}
\includegraphics[angle=0,width=5.5cm]{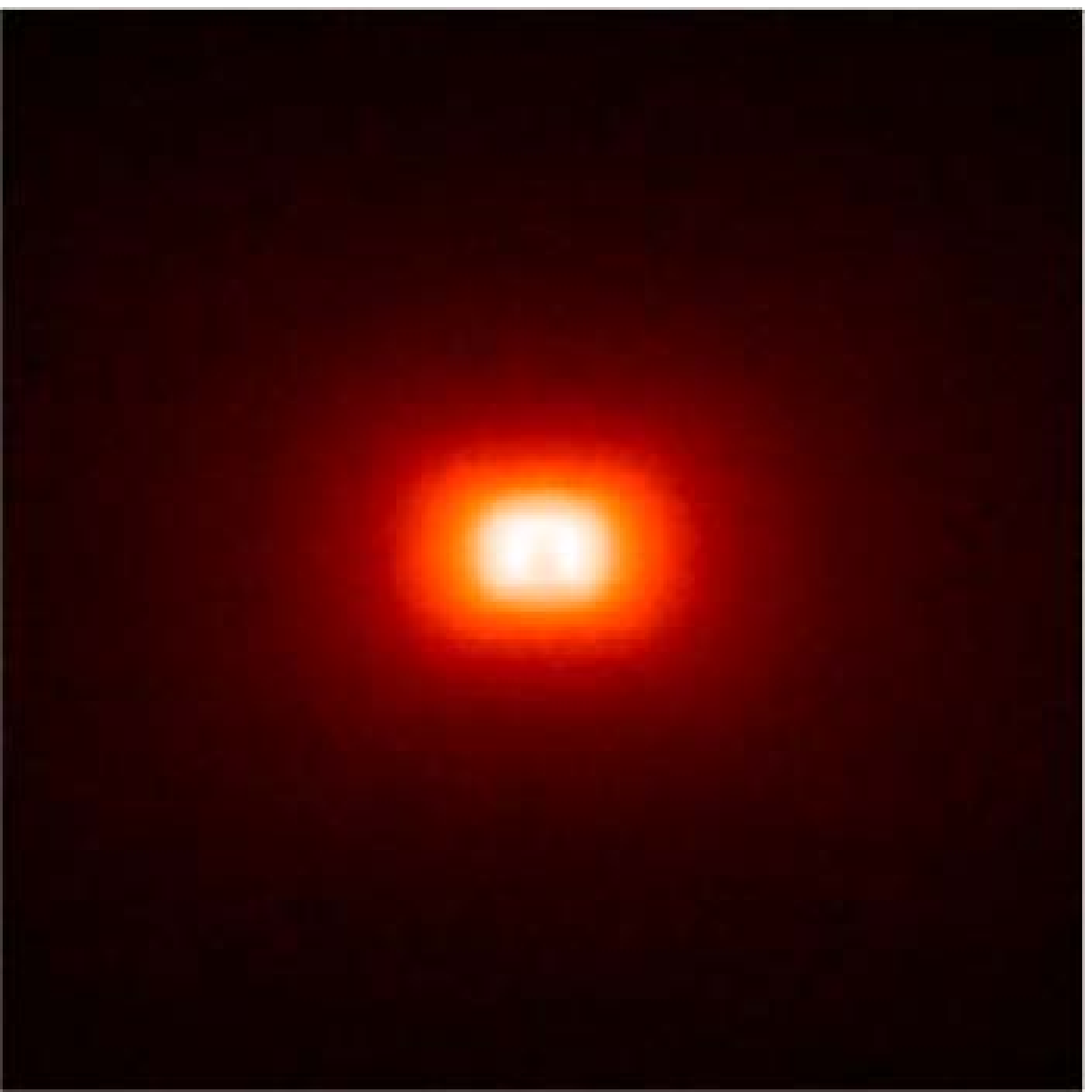}
\end{center}
\caption{NGC\,1068 model images of our clumpy torus model described in Table~\ref{tabpar1068}. From left to right: $K$-, $M$-, and $N$-band images for an inclination angle of $i=55\degr$. The upper row shows images obtained for one particular random cloud arrangement, whereas the images in the lower row are averages of over $\sim$200 different random cloud arrangements. The {\it grainy structure} in the images in the upper row is caused by the {\it clumpiness} of the torus.}\label{figImN1068}
\end{figure*}

In addition to the speckle and MIDI observations, it was possible to observe NGC\,1068 with the VLTI VINCI instrument in the $K$-band \citep{Wit04}. The obtained visibility of $\sim$0.4 at a projected baseline of 46 m corresponds to a structure that is smaller than 3 mas or $\la$$0.2\,{\rm pc}$; i.e., at least $\sim10$ times
smaller than the resolved $\sim$30 mas core. The large visibility of 0.4 at 46 m baseline can be explained by at least two different types of structures within the $18\times39$ mas core: (1) a single compact object (in addition to the $18\times39$ mas core) with a diameter in the range of 0 and 3 mas, or (2) several objects with diameters in the range of 0 and 3 mas. Since there is only one visibility point currently available in the $K$ band at long baselines, it is not yet possible to decide whether this substructure consists of one or several $\la$$0.2$~pc objects. One possible explanation for this substructure is {\it graininess} due to clumpiness, as seen in the images in Figs.~\ref{figImi} and \ref{figImN1068}. The corresponding $K$-band visibilities are shown in Fig.~\ref{visk}. Another possible explanation is that we see the central accretion flow viewed through only moderate extinction directly \citep{Wit04}.

\begin{table}
\caption{Comparison between the observed interferometric sizes and clumpy torus model sizes.}\label{N1068size}
\centering
\vspace{0.1cm}
\begin{tabular}{c c c c}
\hline\hline
Band & model size & interf. size & Reference \\ \hline
$H$ & 2.1$\pm$0.5 pc & 1.3 $\times$ 3.1 pc & W04 \\
$K$ & 2.0$\pm$0.3 pc & 1.3 $\times$ 2.7 pc & W04 \\
$N$ & 2.7$\pm$1.0 pc & $<2.1$ $\times$ 3.4 pc$\,^1$ & J04 \\ \hline
\multicolumn{4}{l}{W04 \citep{Wei04}, J04 \citep{Jaf04}} \\
\multicolumn{4}{l}{$^1$ warm component: 2.1 $\times$ 3.4 pc; hot component: $<1\,{\rm pc}$}
\end{tabular}
\end{table}

Fig.~\ref{figImN1068} presents model images of our clumpy torus model of NGC\,1068 (see Table~\ref{tabpar1068}). These model images are calculated for the wavelength ranges $2.1-2.3\,{\rm\mu m}$, $4.7-5.0\,{\rm\mu m}$, and $9.5-10.3\,{\rm\mu m}$, which approximately correspond to $K$-, $M$-, and $N$-bands, respectively. The images in the upper row are obtained for one particular random cloud arrangement, whereas the images in the lower row are averages over $\sim$200 different random cloud arrangements. The {\it grainy structure} in the images in the upper row is caused by the {\it clumpiness} of the torus and is characteristic for the clumpy torus structure.

Finally, it is possible to compare the resolved $H$-, $K$-, and $N$-band sizes of the compact core of NGC\,1068 with the size predictions of our model described in Table~\ref{tabpar1068}. From the model images, we derived FWHM diameters (Gaussian model fits) of $2.1$, $2.0$, and $2.7\,{\rm pc}$ for the $H$-, $K$-, and $N$-band model diameters. The comparison (Table~\ref{N1068size}) with the interferometrically observed $H$-, $K$-, and $N$-band diameters ($1.3\times 3.1\,{\rm pc}$, $1.3\times 2.8\,{\rm pc}$, and $<2.1\times 3.4\,{\rm pc}$, respectively) shows that the observed and modeled diameters are roughly in agreement. In other words, our model allows us to simultaneously reproduce {\it both} the observed SED and the $H$-, $K$-, and $N$-band diameters of NGC\,1068.

\section{Summary}

We presented a new two-step method for 3-dimensional radiative transfer modeling of {\it clumpy} dust tori of AGN. First, SEDs of individual dust clouds are simulated with our Monte Carlo code. Then, the modeled clouds are randomly arranged within the torus region. The cloud properties are derived from a physically motivated accretion scenario proposed by \citet{Vol04} and \citet{Bec04}. We investigated the influence of clumpy dust distributions on model SEDs and images and presented model calculations for NGC\,1068.

For type 2 objects, we find a moderately pronounced absorption feature which is in agreement with observations. Type 1 model SEDs show moderate emission. The modeled silicate emission-to-continuum flux ratio of $\sim2$ (see Figs. \ref{figVar}, \ref{figCD}, and \ref{figSH}) agrees with recently observed emission features in QSOs and a LINER where the ration is 1.5--2.5 \citep{Sie05,Hao05,Stu05}. In addition, the total NIR and MIR model fluxes are approximately in agreement with observations. It is not necessary to assume additional dust in the polar region to reproduce the observed fluxes.

The actual random cloud arrangement has a significant effect on the SED and images. For the same set of large-scale torus parameters, we find remarkably different SEDs for different random cloud arrangements. The depth of the silicate feature decreases if the number of unobscured clouds is increased in comparison to other random arrangements. Strong statistical variations of the optical depth of the torus due to the clumpiness have already been predicted by \citet{Nen02}. They argue that due to Poisson statistics, clumpiness is able to account for the high variation in X-ray column densities in Seyfert 2 galaxies. We show that clumpiness not only affects absorption of radiation from the central source but also has strong consequences for the IR emission of the torus itself.

For type 1 AGN, we find that the silicate emission may vanish if the toroidal geometry has strong flaring or a high cloud density towards the center. This may apply to Seyfert 1 nuclei with no emission feature observed. In such cases, the absence of a silicate feature does not necessarily mean that there is no dust torus present.

In Sect. \ref{ModSEDN1068} we presented our modeling of the nucleus of NGC\,1068. We showed that our clumpy torus model allows us to roughly reproduce both NGC\,1068's SED and visibilities in the NIR and MIR simultaneously. The obtained strong graininess in the images is caused by the clumpiness. We also showed that clumpy torus models are able to reproduce deeper silicate absorption features with growing interferometric baseline, as reported by \citet{Jaf04}.

\begin{acknowledgements}
We would like to thank Moshe Elitzur for helpful discussions, and Walter Jaffe for providing the correlated MIDI fluxes of NGC\,1068.
\end{acknowledgements}

\appendix

\section{Discussion and assessments of the method}\label{methdis}

As described in \ref{GeomPec}, our treatment of the SOC heating assumes that we have the same number of FOCs around a SOC at all distances from the AGN. In reality, the number of FOCs, however, decreases with growing distance. As a result, we overestimate the ambient radiation field that heats the SOCs. We discuss the influence of our treatment of the ambient radiation field on the torus simulations in the following sections.

\subsection{Energy conservation}\label{energycons}

One important test of our simulations is the study of energy conservation. It is possible to derive a theoretically expected luminosity of the torus. In our model, the average number of clouds along a LOS, $\left<N\right>$, is given by a Gaussian distribution:
\begin{equation}
\left<N\right> = N_0 \cdot \exp\left(-\frac{1}{2}\frac{z^2}{H^2}\right) = N_0 \cdot \exp\left(-\frac{1}{2}\frac{r^2\tan^2\alpha}{H^2}\right) \label{ps1}
\end{equation}
where $\alpha$ is the latitudal angle in spherical coordinates ($z=r\cdot\tan\alpha$) and $H = H(r)$ is the scale height. For simplicity, we assume $H(r) = h \cdot \frac{r}{r_{\rm subl}}$ in the following discussion. $\left<N\right>$ is the expectation value for Poisson statistics \citep{Nen02}. The probability of finding $n$ clouds along a LOS is given by
\begin{equation}
P(n)=\frac{\left<N\right>^n}{n!} \exp\left(-\left<N\right>\right). \label{ps2}
\end{equation}
The probability that we see the AGN unobscured along a LOS is obtained for $n$=0:
\begin{equation}
P(0) = \exp\left(-\left<N\right>\right) = \exp\left(-N_0 \cdot \exp\left(-\frac{1}{2}\frac{r_{\rm subl}^2\tan^2\alpha}{h^2}\right)\right) \label{ps3}
\end{equation}
It is our aim to find the probability that a photon emitted by the AGN in a random direction is hitting a cloud in the torus. For this goal, we simply determine from eq. (\ref{ps3}) the probability $w_0$ that a photon emitted by the AGN in a random direction is escaping the AGN undisturbed by averaging all possible LOS:
\begin{equation}
w_0 = \frac{2}{\pi} \int_0^{2\pi}P(0)\,{\rm d}\alpha \nonumber
\end{equation}
The probability $w_T$ that a photon hits a cloud in the torus is given by
\begin{equation}
w_T = 1-w_0 \nonumber
\end{equation}
Thus, the theoretically expected luminosity $L_{\rm theor}$ of the torus is
\begin{equation}
L_{\rm theor} = w_T \cdot L_{\rm AGN} = (1-w_0) \cdot L_{\rm AGN} \nonumber
\end{equation}
This is the theoretical reference luminosity for our simulations. For energy conservation in the Monte Carlo simulations of the clouds, $L_{\rm theor}$ should be equal to the (unobscured) re-emitted luminosity $L_{\rm FOC}$ of all first order clouds in the torus. We made a number of simulations which showed that in our simulations $L_{\rm theor} = L_{\rm FOC}$ to an accuracy of $\sim$2\%.

Finally, this luminosity $L_{\rm FOC}$ should be re-emitted by our simulated tori when obscuration and SOCs (ambient radiation field) are considered in the simulations, so that $L_{\rm theor} = L_{\rm FOC} = L_{\rm sim}$. Our studies showed that in our simulations
$$L_{\rm theor} = L_{\rm FOC} \pm \sim2\% = L_{\rm sim} \pm \sim5\%.$$
Differences of the order of $\sim$5\% are observed, depending on the actual cloud distribution function. From this analysis we conclude that energy is conserved in our
simulations with satisfying accuracy.

\subsection{SED shape}

We have carefully investigated the effect of our SOC approximation on the total torus SED shape. Our SOCs are generally too hot since the heating by surrounding
clouds is overestimated (see Sect.~\ref{GeomPec}). The ambient radiation field around a SOC in the torus is determined by considering the fraction of the sky around a SOC that is covered by FOCs. This fraction is derived within the inner 2.5 sublimation radii, where the heating of SOCs by surrounding FOCs is strongest within the torus. This treatment results in an overestimation of the ambient radiation field around SOCs at larger distances from the AGN.

We investigated at what wavelengths our SOCs contribute to the torus SED. By removing the SOC emission from our simulations, we found out that the SOC contribution is relevant only longward of $\sim20\,{\rm\mu m}$. In the NIR and MIR up to $20\,{\rm\mu m}$, there is no difference in the torus SED simulations with and without SOCs. The discrepancy between a torus simulation with SOCs and without SOCs becomes a factor of $\sim$2.5 only longward of $100\,{\rm\mu m}$. This factor is constant for wavelengths larger than $100\,{\rm\mu m}$.

In a next step, we investigated the influence of the FOCs on the torus SED. In our current treatment, we only consider direct heating by the AGN. FOCs, however, are also heated by surrounding clouds. This missing heating of FOCs mainly concerns the cold backside of the FOCs. To get an estimation of this additional heating, we added a SOC SED from our SOC treatment (i.e., an overestimated SOC SED) to the FOC backside SED. We found out that in our simulations the contribution from additionally heated FOC backsides is small at all wavelengths. In the NIR and MIR, no difference in the SED is seen at all. In the FIR, a small increase in flux can be observed. The increase at $100\,{\rm\mu m}$ is $\sim$10\%. Thus, the contribution of heated FOC backsides can be considered as insignificant in our simulations.

The SEDs obtained with our SOC treatment represent an upper limit for the real SED. In addition, our simulations without SOCs set a lower limit for the SED. The real SED is somewhere in-between. From the discussion above, we can set upper limits for the maximum overestimation and conclude:
\begin{itemize}
\item In the short wavelength range up to $20\,{\rm\mu m}$, our SOC treatment has almost no influence on the SED (i.e., error $<$2\%).
\item Between $20\,{\rm\mu m}$ and $30\,{\rm\mu m}$, our SOC treatment can cause maximum deviations from the real SED of the same order as the variations by different random cloud arrangements (i.e, $\sim$5-10\%).
\item Longward of $100\,{\rm\mu m}$, the flux overestimation due to our SOC treatment is less than a factor of $\sim$2.5.
\item The influence of our SOC treatment on the luminosity is less than 5\% (see Sect.~\ref{energycons}).
\item The heating of FOC backsides by surrounding clouds is insignificant for NIR and MIR, and is only $\sim$10\% at $100\,{\rm\mu m}$.
\end{itemize}
The studies also highlight that the hot front sides of the FOCs are the dominant contributors to the ambient radiation field that heats the SOCs. Therefore, we conclude that our simulations are satisfying shortward of $20\,{\rm\mu m}$. An improvement of the ambient radiation field for longer wavelengths by simulating the local FOC-to-SOC-relation and the $r$-dependence of the FOC density for any cloud distribution will be the topic of a future paper.

\bibliographystyle{aa}

\end{document}